# Solution of the Skyrme-Hartree-Fock-Bogolyubov equations in the Cartesian deformed harmonic-oscillator basis.
# (VI) HFODD (v2.38j): a new version of the program.

J. Dobaczewski,[a,b,1] W. Satuła,[a] B.G. Carlsson,[b] J. Engel,[c] P. Olbratowski,[a] P. Powałowski,[d] M. Sadziak,[d] J. Sarich,[e] N. Schunck,[f,g] A. Staszczak,[f,g,h] M. Stoitsov,[f,g,i] M. Zalewski[a] and H. Zduńczuk[a]

[a]*Institute of Theoretical Physics, Warsaw University ul. Hoża 69, PL-00681 Warsaw, Poland*
[b]*Department of Physics, P.O. Box 35 (YFL), FI-40014 University of Jyväskylä, Finland*
[c]*Department of Physics and Astronomy, CB3255, University of North Carolina, Chapel Hill, NC 27599-3255, USA*
[d]*Department of Physics, Warsaw University, ul. Hoża 69, 00-681, Warsaw, Poland*
[e]*Argonne National Laboratory, 9700 South Cass Avenue, Building 221, Argonne, IL 60439-4844, USA*
[f]*Department of Physics and Astronomy, University of Tennessee, Knoxville, TN 37996, USA*
[g]*Oak Ridge National Laboratory, P.O. Box 2008, Oak Ridge, TN 37831, USA*
[h]*Department of Theoretical Physics, Maria Curie-Skłodowska University, pl. M. Curie-Skłodowskiej 1, 20-031 Lublin, Poland*
[i]*Institute of Nuclear Research and Nuclear Energy, Bulgarian Academy of Sciences, Sofia, Bulgaria*

**Abstract**

We describe the new version (v2.38j) of the code HFODD which solves the nuclear Skyrme-Hartree-Fock or Skyrme-Hartree-Fock-Bogolyubov problem by using the Cartesian deformed harmonic-oscillator basis. In the new version, we have implemented: (i) projection on good angular momentum (for the Hartree-Fock states), (ii) calculation of the GCM kernels, (iii) calculation of matrix elements of the Yukawa interaction, (iv) the BCS solutions for state-dependent pairing gaps, (v) the HFB solutions for broken simplex symmetry, (vi) calculation of Bohr deformation parameters, (vii) constraints on the Schiff moments and scalar multipole moments, (viii) the $D_{2h}^T$ transformations and rotations of wave functions, (ix) quasiparticle blocking for the HFB solutions in odd and odd-odd nuclei, (x) the Broyden method to accelerate the convergence, (xi) the Lipkin-Nogami method to treat pairing correlations, (xii) the exact Coulomb exchange term, (xiii) several utility options, and we have corrected two insignificant errors.

## NEW VERSION PROGRAM SUMMARY

*Title of the program:* HFODD (v2.38j)

[1]E-mail: jacek.dobaczewski@fuw.edu.pl

*Nature of physical problem*
The nuclear mean-field and an analysis of its symmetries in realistic cases are the main ingredients of a description of nuclear states. Within the Local Density Approximation, or for a zero-range velocity-dependent Skyrme interaction, the nuclear mean-field is local and velocity dependent. The locality allows for an effective and fast solution of the self-consistent Hartree-Fock equations, even for heavy nuclei, and for various nucleonic ($n$-particle $n$-hole) configurations, deformations, excitation energies, or angular momenta. Similar Local Density Approximation in the particle-particle channel, which is equivalent to using a zero-range interaction, allows for a simple implementation of pairing effects within the Hartree-Fock-Bogolyubov method.

*Method of solution*
The program uses the Cartesian harmonic oscillator basis to expand single-particle or single-quasiparticle wave functions of neutrons and protons interacting by means of the Skyrme effective interaction and zero-range pairing interaction. The expansion coefficients are determined by the iterative diagonalization of the mean field Hamiltonians or Routhians which depend non-linearly on the local neutron and proton densities. Suitable constraints are used to obtain states corresponding to a given configuration, deformation or angular momentum. The method of solution has been presented in: J. Dobaczewski and J. Dudek, Comput. Phys. Commun. **102** (1997) 166.

*Summary of revisions*

1. Projection on good angular momentum (for the Hartree-Fock states) has been implemented.
2. Calculation of the GCM kernels has been implemented.
3. Calculation of matrix elements of the Yukawa interaction has been implemented.
4. The BCS solutions for state-dependent pairing gaps have been implemented.
5. The HFB solutions for broken simplex symmetry have been implemented.
6. Calculation of Bohr deformation parameters has been implemented.
7. Constraints on the Schiff moments and scalar multipole moments have been implemented.
8. The $D_{2h}^T$ transformations and rotations of wave functions have been implemented.
9. The quasiparticle blocking for the HFB solutions in odd and odd-odd nuclei has been implemented.
10. The Broyden method to accelerate the convergence has been implemented.
11. The Lipkin-Nogami method to treat pairing correlations has been implemented.
12. The exact Coulomb exchange term has been implemented.
13. Several utility options have been implemented.
14. Two insignificant errors have been corrected.

*Restrictions on the complexity of the problem*
The main restriction is the CPU time required for calculations of heavy deformed nuclei and for a given precision required.

*Typical running time*
One Hartree-Fock iteration for the superdeformed, rotating, parity conserving state of $^{152}_{66}\text{Dy}_{86}$ takes about six seconds on the AMD-Athlon 1600+ processor. Starting from the Woods-Saxon wave functions, about fifty iterations are required to obtain the energy converged within the precision of about 0.1 keV. In case when every value of the angular velocity is converged separately, the complete superdeformed band with precisely determined dynamical moments $\mathcal{J}^{(2)}$ can be obtained within forty minutes of CPU on the AMD-Athlon 1600+ processor. This time can be often reduced by a factor of three when a self-consistent solution for a given rotational frequency is used as a starting point for a neighboring rotational frequency.

*Unusual features of the program*
The user must have access to (i) an implementation of the BLAS (Basic Linear Algebra Subroutines), (ii) the NAGLIB subroutine F02AXE, or LAPACK subroutines ZHPEV, ZHPEVX, or ZHEEVR, which diagonalize complex hermitian matrices, and (iii) the LINPACK subroutines ZGEDI and ZGECO, which invert arbitrary complex matrices and calculate determinants, or provide another set of subroutines that can perform such a tasks. The LAPACK and LINPACK subroutines and an unoptimized version of the BLAS can be obtained from the Netlib Repository at the University of Tennessee, Knoxville: http://www.netlib.org/.

## LONG WRITE-UP

# 1 Introduction

The method of solving the Hartree-Fock (HF) equations in the Cartesian harmonic oscillator (HO) basis was described in the previous publication, Ref. [1] (I). Four versions of the code HFODD were previously published: (v1.60r) [2] (II), (v1.75r) [3] (III), (v2.08i) [4] (IV), and (v2.08k) [5] (V). The User's Guide for version (v2.08k) is available in Ref. [6] and the code home page is at http://www.fuw.edu.pl/~dobaczew/hfodd/hfodd.html. The present paper is a long write-up of the new version (v2.38j) of the code HFODD. This extended version features the angular-momentum projection, calculations of the generator-coordinate-method (GCM) kernels, and several other major modifications, and is fully compatible with all previous versions.

Information provided in previous publications [1]–[5] remains valid, unless explicitly mentioned in the present long write-up. Below we refer by Roman numbers (I)–(V) to section and equation numbers in these previous publications

In Section 2 we review modifications introduced in version (v2.38j) of the code HFODD. Section 3 lists all additional new input keywords and data values, introduced in version (v2.38j). The structure of the input data file remains the same as in the previous versions, see Section II-3. Similarly, all previously introduced keywords and data values retain their validity and meaning.

# 2 Modifications introduced in version (v2.38j)

## 2.1 Projection on good angular momentum and calculation of the GCM kernels

The code HFODD (v2.38j) contains a new option allowing for the angular-momentum projection (AMP) after variation of an arbitrary symmetry-unrestricted Slater determinant $|\Phi\rangle$ provided by the code. This includes projection of cranked time-reversal symmetry breaking HF states which is of particular interest in high-spin applications [7, 8, 9, 10, 11, 12].

The angular-momentum conserving wave function is obtained from the Slater determinant $|\Phi\rangle$ by applying the standard SO(3) AMP operator [13, 14]:

$$|IMK\rangle = \hat{P}^I_{MK}|\Phi\rangle = \frac{2I+1}{8\pi^2} \int D^{I*}_{MK}(\Omega)\, \hat{R}(\Omega)|\Phi\rangle\, d\Omega, \qquad (1)$$

projecting onto angular momentum $I$ with projection $M$ and $K$ along the laboratory and intrinsic $z$-axes, respectively. The symbol $\Omega$ labels here a set of three Euler angles $(\alpha, \beta, \gamma)$, $D^I_{MK}(\Omega)$ is the Wigner function and $\hat{R}(\Omega) = e^{-i\alpha \hat{I}_z} e^{-i\beta \hat{I}_y} e^{-i\gamma \hat{I}_z}$ is the active (body) rotation operator [15].

The intrinsic quantum number $K$ is, in general, not conserved. The $K$-mixing is taken into account by assuming the following form for the eigenstates [14]:

$$|i; IM\rangle = \sum_K g_{IK}^{(i)} |IMK\rangle \equiv \sum_K g_{IK}^{(i)} \hat{P}_{MK}^I |\Phi\rangle. \quad (2)$$

The mixing coefficients $g_{IK}^{(i)}$ are determined by a minimization of energy, which amounts to solving the following Hill-Wheeler (H-W) eigenvalue problem:

$$\sum_{K'} H_{KK'} g_{IK'}^{(i)} = E_i \sum_{K'} N_{KK'} g_{IK'}^{(i)}, \quad (3)$$

separately for each angular momentum $I$. The Hamiltonian and norm matrix elements entering the H-W equation (3) equal:

$$H_{KK'} = \langle \Phi | \hat{H} \hat{P}_{KK'}^I | \Phi \rangle = \frac{2I+1}{8\pi^2} \int d\Omega \, D_{KK'}^{I*}(\Omega) \, \mathcal{H}(\Omega), \quad (4)$$

$$N_{KK'} = \langle \Phi | \hat{P}_{KK'}^I | \Phi \rangle = \frac{2I+1}{8\pi^2} \int d\Omega \, D_{KK'}^{I*}(\Omega) \, \mathcal{N}(\Omega), \quad (5)$$

where

$$\mathcal{H}(\Omega) = \langle \Phi | \hat{H} \hat{R}(\Omega) | \Phi \rangle, \quad (6)$$
$$\mathcal{N}(\Omega) = \langle \Phi | \hat{R}(\Omega) | \Phi \rangle, \quad (7)$$

denote the Hamiltonian and norm kernels, respectively.

The H-W eigenvalue problem (3) should be handled with care due to its overcompleteness. This difficulty is overcome in the code by solving the problem (3) in the *collective basis* spanned by the *natural states*:

$$|IM\rangle^{(m)} = \frac{1}{\sqrt{n_m}} \sum_K \bar{\eta}_K^{(m)} |IMK\rangle. \quad (8)$$

These states are constructed of the eigenstates of the norm matrix:

$$\sum_{K'} N_{KK'} \bar{\eta}_{K'}^{(m)} = n_m \, \bar{\eta}_K^{(m)}, \quad (9)$$

having non-zero $n_m > 0$ eigenvalues. More precisely, due to numerical stability reasons, the *collective subspace* is constructed by using only $n = 1, 2, ..., m_{\max}$ eigenstates having $n_m > \zeta$, where $\zeta$ is an externally provided basis cut-off parameter. Final diagonalization of the Hamiltonian matrix is performed in the $m_{\max}$-dimensional *collective subspace* defined as:

$$|i; IM\rangle = \sum_{m=1}^{m_{\max}} f_m^{(i)} |IM\rangle^{(m)}. \quad (10)$$

The code provides, for each angular momentum, eigenenergies and mixing coefficients recalculated to the representation defined by Eq. (2) according to the following formula:

$$g_{IK}^{(i)} = \sum_{m=1}^{m_{\max}} \frac{f_m^{(i)} \eta_K^{(m)}}{\sqrt{n_m}}. \quad (11)$$

The cornerstone of the AMP scheme described above is a calculation of the Hamiltonian (6) and norm (7) kernels. A prerequisite for this calculation is a spatial rotation of the Slater determinant. It amounts to rotating independently each single-particle state $\hat{R}(\Omega)\,\varphi_i(\vec{r},\sigma) \equiv \tilde{\varphi}_i(\vec{r},\sigma)$:

$$\hat{R}(\Omega)|\Phi\rangle \equiv |\tilde{\Phi}\rangle = \frac{1}{\sqrt{A!}} \begin{vmatrix} \tilde{\varphi}_1(1) & \tilde{\varphi}_2(1) & \ldots & \tilde{\varphi}_A(1) \\ \tilde{\varphi}_1(2) & \tilde{\varphi}_2(2) & \ldots & \tilde{\varphi}_A(2) \\ \vdots & \vdots & \ddots & \vdots \\ \tilde{\varphi}_1(A) & \tilde{\varphi}_2(A) & \ldots & \tilde{\varphi}_A(A) \end{vmatrix}. \tag{12}$$

This particular task is performed by the subroutine ROTWAV. Taking advantage of a fact that the HFODD is coded in the Cartesian HO basis[2] $\psi_{n_x}(x)\psi_{n_y}(y)\psi_{n_z}(z)$, cf. Eq. (I-71), the three-dimensional rotation is factorized into three independent and numerically equivalent one-dimensional rotations around $z-$, $y-$ and again $z-$axis, respectively. To be more specific, in the first step, each single-particle wave function is rotated around the $z$-axis by the Euler angle $\gamma$, and the resulting rotated wave function is expanded in the original Cartesian harmonic oscillator basis:

$$\begin{aligned}
\hat{R}_z(\gamma)\varphi_i(\vec{r},\sigma) &= \hat{R}_z(\gamma) \sum_{n_x n_y n_z, s_z} A_i^{n_x n_y n_z, s_z} \psi_{n_x}(x)\,\psi_{n_y}(y)\,\psi_{n_z}(z)\,\chi_{s_z}(\sigma) \\
&= \sum_{n_x n_y n_z, s_z} \tilde{A}_i^{n_x n_y n_z, s_z}(\gamma)\,\psi_{n_x}(x)\,\psi_{n_y}(y)\,\psi_{n_z}(z)\,\chi_{s_z}(\sigma).
\end{aligned} \tag{13}$$

The expansion coefficients $\tilde{A}_i^{n_x n_y n_z, s_z}(\gamma)$ are calculated by using the one-dimensional rotation of the basis vectors:

$$e^{-i\gamma \hat{L}_z} \psi_{n_x}(x)\,\psi_{n_y}(y) = \sum_{n'_x n'_y} B_{n_x n_y}^{n'_x n'_y}(\gamma) \psi_{n'_x}(x)\,\psi_{n'_y}(y), \tag{14}$$

and the coefficients $B_{n_x n_y}^{n'_x n'_y}(\gamma)$, which are non-zero only for $n_x + n_y = n'_x + n'_y$, are calculated numerically by using the Gauss-Hermite (G-H) quadratures. Rotations around the $y-$ and $z-$axes by the Euler angles $\beta$ and $\alpha$, respectively, are performed in exactly the same way. This method is exact, numerically very efficient, and inherent to the HFODD code. Indeed, since the rotated state is expanded in the original HO basis the structure of the code is preserved and allows, after only a minor generalization, for using the existing subroutines of the code to calculate the Hamiltonian $\mathcal{H}(\Omega)$ and norm $\mathcal{N}(\Omega)$ kernels.

Calculation of the kernel, $\langle\Phi|\hat{\mathcal{O}}|\tilde{\Phi}\rangle$, for an arbitrary quantum operator $\hat{\mathcal{O}}$, proceeds along standard rules for calculating matrix elements between two non-orthogonal Slater determinants, see e.g. [16]. In particular, the norm kernel is given by:

$$\langle\Phi|\tilde{\Phi}\rangle = \text{Det}\,\bar{O}, \tag{15}$$

with the overlap matrix elements defined as:

$$O_{ij} = \int d\vec{r} \sum_\sigma \tilde{\varphi}_i(\vec{r},\sigma) \varphi_j^*(\vec{r},\sigma). \tag{16}$$

---

[2]The present implementation assumes a spherical HO basis to ensure that the rotation does not induce components of single-particle wave functions that would go beyond the assumed basis.

The kernel of an arbitrary one-body operator $\hat{F}$ reads:

$$\frac{\langle \Phi|\hat{F}|\tilde{\Phi}\rangle}{\langle \Phi|\tilde{\Phi}\rangle} = \sum_{ij}\langle \varphi_j|\hat{F}|\tilde{\varphi}_i\rangle O_{ij}^{-1} = \iint \mathrm{d}\vec{r}\mathrm{d}\vec{r}' \sum_{\sigma\sigma'} \langle \vec{r}\sigma|\hat{F}|\vec{r}'\sigma'\rangle \, \tilde{\rho}(\vec{r}'\sigma',\vec{r}\sigma), \qquad (17)$$

where $O_{ij}^{-1}$ denotes the matrix element of the inverse of the overlap matrix $\bar{O}$ while $\tilde{\rho}(\vec{r}'\sigma',\vec{r}\sigma)$ is the one-body transition density matrix defined as:

$$\tilde{\rho}(\vec{r}'\sigma',\vec{r}\sigma) = \sum_{ij} \varphi_i^*(\vec{r},\sigma)\, \tilde{\varphi}_j(\vec{r}',\sigma')\, O_{ji}^{-1}. \qquad (18)$$

At the same time, the most general GCM kernels can be calculated by using two different sets of wave functions for $\varphi_i(\vec{r},\sigma)$ and $\tilde{\varphi}_j(\vec{r},\sigma)$, which correspond to two arbitrary different Slater determinants.

The kernel (17) and density matrix (18) can be further simplified by introducing auxiliary ket-states defined as:

$$\tilde{\phi}_i(\vec{r}'\sigma') \equiv \sum_j \tilde{\varphi}_j(\vec{r}'\sigma')\, O_{ji}^{-1}. \qquad (19)$$

Indeed, this allows for rewriting the transition density matrix to a "diagonal" form:

$$\tilde{\rho}(\vec{r},\sigma,\vec{r}',\sigma') \equiv \sum_i \varphi_i^*(\vec{r},\sigma)\, \tilde{\phi}_i(\vec{r}',\sigma'), \qquad (20)$$

where the summation goes over a single index in full analogy to the diagonal HF density matrix. It means that the transition density matrix and one-body kernels can be calculated by using the standard subroutines of the code. All what needs to be done is a substitution of the HF ket-state by the auxiliary ket-state (19).

Similar property also holds for the two-body-interaction kernel, which preserves the functional form of an energy density functional (EDF) derived for this interaction by averaging it over the Slater determinant. Again, all what needs to be done is a replacement of the density matrix by the transition density matrix. In our particular case, the Skyrme EDF can be expressed by using six local isoscalar ($t=0$) and six local isovector ($t=1$) densities, including the particle $\rho_t$, kinetic $\tau_t$, spin $\vec{s}_t$, spin-kinetic $\vec{T}_t$, current $\vec{j}_t$, and spin-current $\overleftrightarrow{J}_t$ densities and their derivatives, see Eqs. (I-5)–(I-7). The Skyrme-interaction kernel preserves the functional form of the Skyrme EDF:

$$\begin{aligned}
\frac{\langle \Phi|V_{Sk}|\tilde{\Phi}\rangle}{\langle \Phi|\tilde{\Phi}\rangle} &\equiv \frac{\mathcal{H}_{Sk}(\Omega)}{\mathcal{N}(\Omega)} = \\
&= \sum_{t=0,1} \int d^3\vec{r} \Big[ C_t^\rho[\tilde{\rho}_0]\tilde{\rho}_t^2 + C_t^{\Delta\rho}\tilde{\rho}_t\Delta\tilde{\rho}_t + C_t^\tau \tilde{\rho}_t\tilde{\tau}_t + C_t^J \overleftrightarrow{\tilde{J}}_t^2 + C_t^{\nabla J}\tilde{\rho}_t\vec{\nabla}\cdot\vec{\tilde{J}}_t \\
&\quad + C_t^s[\tilde{\rho}_0]\vec{\tilde{s}}_t^2 + C_t^{\Delta s}\vec{\tilde{s}}_t\cdot\Delta\vec{\tilde{s}}_t + C_t^T \vec{\tilde{s}}_t\cdot\vec{\tilde{T}}_t + C_t^j \vec{\tilde{j}}_t^2 + C_t^{\nabla j}\vec{\tilde{s}}_t\cdot\left(\vec{\nabla}\times\vec{\tilde{j}}_t\right)\Big].
\end{aligned} \qquad (21)$$

It depends on six local transition densities $\tilde{\rho}_t, \tilde{\tau}_t, \vec{\tilde{s}}_t, \vec{\tilde{T}}_t, \vec{\tilde{j}}_t, \overleftrightarrow{\tilde{J}}_t$, which are the counterparts of the six diagonal local densities mentioned above.

However, in the case of density-dependent interactions, like the Skyrme force, the kernel is not fully defined. The procedure must be augmented by a prescription concerning the treatment of the density-dependent term. In the HFODD code we implement the standard prescription, that is, in the primary coupling constants we replace the isoscalar density by its transition counterpart, i.e., $C_t^\rho[\rho_0] \to C_t^\rho[\tilde{\rho}_0]$ and $C_t^s[\rho_0] \to C_t^s[\tilde{\rho}_0]$ (see discussion in Ref. [17]).

As already mentioned, implementation of the AMP requires a relatively minor recoding of the standard routines of the code. This concerns two generic matrices, $D$ and $L$ (I-39), which are used in the HFODD code to encode density matrices and their derivatives. In version (v2.38j), these matrices were generalized in the following natural way:

$$D_{\hat{\mu}\hat{\nu},\alpha}^{qq'} = \sum_i \left(\nabla_{\hat{\mu}}\tilde{\phi}_i(\vec{r}\sigma)\right)\left(\nabla_{\hat{\nu}}\varphi_i^*(\vec{r}\sigma')\right), \tag{22}$$

and

$$L^{qq'} = \frac{1}{2}\sum_i \left(\tilde{\phi}_i(\vec{r}\sigma)\Delta\varphi_i^*(\vec{r}\sigma') + \Delta\tilde{\phi}_i(\vec{r}\sigma)\varphi_i^*(\vec{r}\sigma')\right), \tag{23}$$

where $\nabla_{\hat{\mu}} \equiv (1, \vec{\nabla})$ for $\hat{\mu} = 0, 1, 2, 3$ and the indices $q$ and $q'$ denote the signs of spins $\sigma$ and $\sigma'$, respectively. In order to allow for non-hermitian, complex transition densities and their derivatives, the set of formulas expressing them in terms of the $D$ and $L$ matrices, see Eqs. (I-40)–(I-51), was generalized in the following way:

- scalar densities

$$\tilde{\rho} = D_{00}^{++} + D_{00}^{--}, \tag{24}$$

$$\tilde{\tau} = \sum_\mu \left(D_{\mu\mu}^{++} + D_{\mu\mu}^{--}\right), \tag{25}$$

$$\Delta\tilde{\rho} = 2\tilde{\tau} + 2\left(L^{++} + L^{--}\right), \tag{26}$$

$$\vec{\nabla}\cdot\vec{\tilde{J}} = i\left(D_{12}^{++} - D_{12}^{--} - D_{21}^{++} + D_{21}^{--}\right) + i\left(D_{23}^{+-} + D_{23}^{-+} - D_{32}^{+-} - D_{32}^{-+}\right) - \left(D_{31}^{+-} - D_{31}^{-+} - D_{13}^{+-} + D_{13}^{-+}\right), \tag{27}$$

- vector densities

$$\tilde{s}_1 = D_{00}^{+-} + D_{00}^{-+}, \tag{28a}$$

$$\tilde{s}_2 = i\left(D_{00}^{+-} - D_{00}^{-+}\right), \tag{29a}$$

$$\tilde{s}_3 = D_{00}^{++} - D_{00}^{--}, \tag{30a}$$

$$\tilde{T}_1 = \sum_\mu \left(D_{\mu\mu}^{+-} + D_{\mu\mu}^{-+}\right), \tag{31a}$$

$$\tilde{T}_2 = i\sum_\mu \left(D_{\mu\mu}^{+-} - D_{\mu\mu}^{-+}\right), \tag{32a}$$

$$\tilde{T}_3 = \sum_\mu \left(D_{\mu\mu}^{++} - D_{\mu\mu}^{--}\right), \tag{33a}$$

$$\Delta \tilde{s}_1 = 2\tilde{T}_1 + 2\left(L^{+-} + L^{-+}\right), \tag{34a}$$

$$\Delta \tilde{s}_2 = 2\tilde{T}_2 + 2i\left(L^{+-} - L^{-+}\right), \tag{35a}$$

$$\Delta \tilde{s}_3 = 2\tilde{T}_3 + 2\left(L^{++} - L^{--}\right), \tag{36a}$$

$$\nabla_\mu \tilde{\rho} = D^{++}_{\mu 0} + D^{--}_{\mu 0} + D^{++}_{0\mu} + D^{--}_{0\mu}, \tag{37}$$

$$\tilde{j}_\mu = \frac{1}{2i} D^{++}_{\mu 0} + D^{--}_{\mu 0} - D^{++}_{0\mu} - D^{--}_{0\mu}, \tag{38}$$

$$\left(\nabla \times \tilde{s}\right)_1 = \left(D^{++}_{02} - D^{--}_{02} + D^{++}_{20} - D^{--}_{20}\right) - i\left(D^{+-}_{03} + D^{+-}_{30} - D^{-+}_{03} - D^{-+}_{30}\right), \tag{39a}$$

$$\left(\nabla \times \tilde{s}\right)_2 = \left(D^{+-}_{03} + D^{-+}_{03} + D^{+-}_{30} + D^{-+}_{30}\right) - \left(D^{++}_{01} - D^{--}_{01} + D^{++}_{10} - D^{--}_{10}\right), \tag{40a}$$

$$\left(\nabla \times \tilde{s}\right)_3 = i\left(D^{+-}_{01} + D^{+-}_{10} - D^{-+}_{01} - D^{-+}_{10}\right) - \left(D^{+-}_{02} + D^{+-}_{20} + D^{-+}_{02} + D^{-+}_{20}\right), \tag{41a}$$

$$\left(\nabla \times \tilde{j}\right)_1 = i\left(D^{++}_{23} - D^{++}_{32} + D^{--}_{23} - D^{--}_{32}\right), \tag{42a}$$

$$\left(\nabla \times \tilde{j}\right)_2 = i\left(D^{++}_{31} - D^{++}_{13} + D^{--}_{31} - D^{--}_{13}\right), \tag{43a}$$

$$\left(\nabla \times \tilde{j}\right)_3 = i\left(D^{++}_{12} - D^{++}_{21} + D^{--}_{12} - D^{--}_{21}\right), \tag{44a}$$

- tensor density

$$\tilde{J}_{\mu 1} = \frac{1}{2i}\left(D^{+-}_{\mu 0} + D^{-+}_{\mu 0} - D^{+-}_{0\mu} - D^{-+}_{0\mu}\right), \tag{45a}$$

$$\tilde{J}_{\mu 2} = \frac{1}{2}\left(D^{+-}_{\mu 0} - D^{-+}_{\mu 0} - D^{+-}_{0\mu} + D^{-+}_{0\mu}\right), \tag{46a}$$

$$\tilde{J}_{\mu 3} = \frac{1}{2i}\left(D^{++}_{\mu 0} - D^{--}_{\mu 0} - D^{++}_{0\mu} + D^{--}_{0\mu}\right). \tag{47a}$$

The final step in the calculation of the Hamiltonian and norm kernels (4)-(5) is the numerical integration over the Euler angles. In our implementation, these three-dimensional integrals are calculated by using the Gauss quadratures [18]. To achieve a maximum accuracy, the Gauss-Tchebyschev (G-T) quadrature is used for the integration over the $\alpha$ and $\gamma$ Euler angles and the Gauss-Legendre (G-L) quadrature is used for the integration over the $\beta$ angle. This combined technique ensures, in fact, exact integration, provided that the numbers of the G-T nodes $n_\alpha = n_\gamma$ and the number of the G-L nodes $n_\beta$ are sufficiently large.

The numbers of the Gauss nodes to be used depend on the value of the maximum-spin component $I_{max}$ in the Slater determinant expansion in good angular-momentum basis states (1):

$$|\Phi\rangle = \sum_{I=I_{min}}^{I_{max}} \sum_{K=-I}^{I} |IKK\rangle. \tag{48}$$

Therefore, condition $2I_{max} \leq n_i - 1$ for $i = \alpha, \beta, \gamma$ is common for all three Euler angles. In practice, twice as large numbers of nodes ensure good numerical stability for terms that are not polynomial, like for example the density-dependent terms.

Note, however, that since $I_{max}$ is *a priori* unknown, the precision of integration can be verified only *a posteriori*. The user can verify this precision by inspecting the completeness relations for the overlap:

$$1 = \sum_{IK} \langle IKK|IKK\rangle \equiv \sum_{IK} \frac{2I+1}{8\pi^2} \int d\Omega D_{KK}^{I*} \langle \Phi|\hat{R}(\Omega)|\Phi\rangle, \qquad (49)$$

and the Hamiltonian (or the Hartree-Fock energy $E_{HF}$):

$$E_{HF} \equiv \langle \Phi|\hat{H}|\Phi\rangle = \sum_{IK} \langle IKK|\hat{H}|IKK\rangle = \sum_{IK} \frac{2I+1}{8\pi^2} \int d\Omega D_{KK}^{I*} \langle \Phi|\hat{H}\hat{R}(\Omega)|\Phi\rangle. \qquad (50)$$

The results concerning the completeness relations constitute part of the standard printout of the code.

## 2.2 Calculation of electromagnetic transition probabilities

Angular momentum projection opens up a possibility to compute fully quantum mechanically transition rates for electromagnetic radiation between final $\langle f; I_f M_f|$ and initial $|i; I_i M_i\rangle$ $K$-mixed states (2) of angular momenta $I_f$ and $I_i$, respectively. Reduced transition probabilities are defined by means of the reduced matrix elements, which are provided by the code, as :

$$\begin{aligned} B(E\lambda, I_i \longrightarrow I_f) &= \frac{1}{2I_i+1} \left| \langle f; I_f||\hat{Q}_\lambda||i; I_i\rangle \right|^2, \\ B(M\lambda, I_i \longrightarrow I_f) &= \frac{1}{2I_i+1} \left| \langle f; I_f||\hat{M}_\lambda||i; I_i\rangle \right|^2, \end{aligned} \qquad (51)$$

for electric $\hat{Q}_{\lambda\mu} = r^\lambda Y_{\lambda\mu}(\varphi, \theta)$ and magnetic $\hat{M}_{\lambda\mu} = g_s \hat{M}_{s;\lambda\mu} + g_l \hat{M}_{l;\lambda\mu}$ transition operators, where $g_s$ and $g_l$ denote spin and orbital gyromagnetic factors, respectively. The spin and orbital parts of the magnetic transition operator equal $\hat{M}_{s;\lambda\mu} = (\vec{\nabla} Q_{\lambda\mu}) \cdot \hat{\vec{S}}$ and $\hat{M}_{l;\lambda\mu} = \frac{2}{\lambda+1}(\vec{\nabla} Q_{\lambda\mu}) \cdot \hat{\vec{L}}$, respectively.

The electric $\hat{Q}_{\lambda\mu}$ and magnetic $\hat{M}_{\lambda\mu}$ transition operators transform under spatial rotations like components of spherical tensor $\hat{T}_{\lambda\mu}$ of rank $\lambda$. According to the Wigner-Eckart theorem, the matrix elements of these operators are, therefore, equal to:

$$\langle f; I_f M_f|\hat{T}_{\lambda\mu}|i; I_i M_i\rangle = (-1)^{2\lambda} C_{I_i M_i \lambda\mu}^{I_f M_f} \frac{\langle f; I_f||\hat{T}_\lambda||i; I_i\rangle}{\sqrt{2I_f+1}}, \qquad (52)$$

where $C_{I_i M_i \lambda\mu}^{I_f M_f}$ denotes the Clebsch-Gordan coefficient.

The left-hand side of Eq. (52) can be calculated by using the projected states (1). Using the definition of the $K$-mixed states (2), it can be written as:

$$\begin{aligned} \langle f; I_f M_f|\hat{T}_{\lambda\mu}|i; I_i M_i\rangle &= \sum_{K_i K_f} g_{I_f K_f}^{(f)*} g_{I_i K_i}^{(i)} \langle I_f M_f K_f|\hat{T}_{\lambda\mu}|I_i M_i K_i\rangle \\ &= \sum_{K_i K_f} g_{I_f K_f}^{(f)*} g_{I_i K_i}^{(i)} \langle \Phi_f|\hat{P}_{K_f M_f}^{I_f} \hat{T}_{\lambda\mu} \hat{P}_{M_i K_i}^{I_i}|\Phi_i\rangle, \end{aligned} \qquad (53)$$

where $|\Phi_f\rangle$ and $|\Phi_i\rangle$ are, in principle, different Slater determinants. The matrix element (53) can be further simplified after applying the following transformation rule, valid for arbitrary spherical tensor:

$$\hat{P}^{I_f}_{K_f M_f} \hat{T}_{\lambda\mu} \hat{P}^{I_i}_{M_i K_i} = C^{I_f M_f}_{I_i M_i \lambda\mu} \sum_{M\mu'} (-1)^{2\mu'} C^{I_f K_f}_{I_i M \lambda\mu'} \, \hat{T}_{\lambda\mu'} \hat{P}^{I_i}_{M K_i}. \tag{54}$$

Indeed, by using this formula one obtains:

$$\langle I_f M_f K_f | \hat{T}_{\lambda\mu} | I_i M_i K_i \rangle = C^{I_f M_f}_{I_i M_i \lambda\mu} \sum_{M\mu'} (-1)^{2\mu'} C^{I_f K_f}_{I_i M \lambda\mu'} \langle \Phi_f | \hat{T}_{\lambda\mu'} P^{I_i}_{M K_i} | \Phi_i \rangle, \tag{55}$$

what brings the calculation of the transition rates down to a standard task of the AMP, namely to the calculation of the matrix elements of the one-body operator $\langle \Phi_f | \hat{T}_{\lambda\mu'} \hat{P}^{I_i}_{M K_i} | \Phi_i \rangle$ which is described in detail in Section 2.1. By comparing Eq. (52) for $f \equiv K_f$ and $i \equiv K_i$ with Eq. (55), we finally have:

$$\langle I_f K_f || \hat{T}_\lambda || I_i K_i \rangle = \sqrt{2I_f + 1} \sum_{M\mu'} C^{I_f K_f}_{I_i M \lambda\mu'} \langle \Phi_f | \hat{T}_{\lambda\mu'} \hat{P}^{I_i}_{M K_i} | \Phi_i \rangle. \tag{56}$$

These quantities enter directly the equation for the reduced matrix elements between the AMP $K$-mixed states, which reads:

$$\langle f; I_f || \hat{T}_\lambda || i; I_i \rangle = \sum_{K_i K_f} g^{(f)*}_{I_f K_f} g^{(i)}_{I_i K_i} \langle I_f K_f || \hat{T}_\lambda || I_i K_i \rangle. \tag{57}$$

For the sake of completeness, it should be recalled that in the code HFODD the multipole operators are defined as $\hat{Q}_{\lambda\mu} \equiv r^\lambda Y^*_{\lambda\mu}$, see Eq. (IV-2). Hence, the matrix elements of multipole electric and magnetic operators are calculated in the code for dual tensors. In order to be consistent with the matrix elements of multipole operators printed by the code one needs, therefore, to take $\hat{T}_{\lambda\mu} = \hat{Q}^*_{\lambda\mu}$ for electric transitions and use $\vec{\nabla} Q^*_{\lambda\mu}$ in the definition of magnetic transitions. Of course, these phase conventions do not influence the reduced transition probabilities (51).

### 2.3 The Yukawa interaction

The code HFODD can now evaluate the expectation value of a two-body Yukawa interaction

$$V = \sum_{i<j} \frac{e^{-\alpha r_{ij}}}{r_{ij}} \left[ a + b\sigma_i \cdot \sigma_j + c\tau_i \cdot \tau_j + d\sigma_i \cdot \sigma_k \tau_i \cdot \tau_k \right],$$

where $r_{ij} \equiv |\mathbf{r}_i - \mathbf{r}_j|$, and $\alpha$, $a, b, c,$ and $d$ are arbitrary constants. The code makes use of an approximate expansion of the Yukawa function in terms of Gaussians:

$$\frac{e^{-x}}{x} \approx 6.79\, e^{-34x^2} + 2.41\, e^{-6.6x^2} + 0.786\, e^{-1.44x^2} + 0.241\, e^{-0.38x^2} - 0.062\, e^{-0.15x^2} + 0.078\, e^{-0.13x^2}, \tag{58}$$

Of course, the Yukawa function is singular at the origin and Gaussians are not, but the volume element contains a factor of $r^2$ that makes the approximation quite accurate near the origin (it is a bit less so at larger distances).

As described in Ref. [19], HFODD can also use a series of Gaussians to approximate the expectation value of the CP-violating potential produced by pion exchange with a CP-odd pion-nucleon vertex:

$$\hat{V}_{PT}(\boldsymbol{r}_1 - \boldsymbol{r}_2) = -\frac{g\, m_\pi^2 \hbar\, c}{8\pi m_N}\left\{(\boldsymbol{\sigma}_1 - \boldsymbol{\sigma}_2)\cdot(\boldsymbol{r}_1 - \boldsymbol{r}_2)\Big[\bar{g}_0\, \vec{\tau}_1\cdot\vec{\tau}_2 - \frac{\bar{g}_1}{2}(\tau_{1z} + \tau_{2z})\right.$$
$$\left. + \bar{g}_2(3\tau_{1z}\tau_{2z} - \vec{\tau}_1\cdot\vec{\tau}_2)\Big] - \frac{\bar{g}_1}{2}(\boldsymbol{\sigma}_1 + \boldsymbol{\sigma}_2)\cdot(\boldsymbol{r}_1 - \boldsymbol{r}_2)(\tau_{1z} - \tau_{2z})\right\}$$
$$\times \frac{\exp(-m_\pi|\boldsymbol{r}_1 - \boldsymbol{r}_2|)}{m_\pi|\boldsymbol{r}_1 - \boldsymbol{r}_2|^2}\left[1 + \frac{1}{m_\pi|\boldsymbol{r}_1 - \boldsymbol{r}_2|}\right], \tag{59}$$

where the $g$'s label the isoscalar, isovector, and isotensor pion-nucleon coupling strengths and $m_\pi$ and $m_N$ stand, respectively, for the pion and nucleon mass in units of fm$^{-1}$. To simulate the effects of short-range correlations, absent from HF wave functions, we include a phenomenological correlation function [20]

$$f(r) = 1 - e^{-1.1 r^2}(1 - 0.68\, r^2)\,, \tag{60}$$

and make the approximation

$$g(r) = f(r)^2\, \frac{e^{-m_\pi r}}{r^2}\left(1 + \frac{1}{m_\pi r}\right) \approx 1.75\, e^{-1.1 r^2} + 0.53\, e^{-0.68 r^2} + 0.11\, e^{-0.21 r^2} + 0.004\, e^{-0.06 r^2}, \tag{61}$$

where $m_\pi \equiv 0.7045$ fm$^{-1}$ and the numbers in the fit all have units of fm$^{-2}$. The extra factor of $\boldsymbol{r}$ not included in Eq. (61) (i.e. the factor $\boldsymbol{r}_1 - \boldsymbol{r}_2$ in Eq. (59)) is treated separately.

A Gaussian interaction between, e.g., particles 1 and 2, leaving out the spin dependence, factors into functions that each depend only on the $x$, $y$, or $z$ component of the vector difference $\boldsymbol{r}_1 - \boldsymbol{r}_2$. To evaluate the expectation value of the first Gaussian, which depends on $x_1 - x_2$, it is easiest to write the product of two oscillator wave functions (of $x_1$ and $x_2$) in terms of relative and center-of-mass coordinates. To do so, we use

$$a^\dagger_{CM} = \frac{a^\dagger_1 + a^\dagger_2}{\sqrt{2}}, \qquad a^\dagger_{\rm rel} = \frac{a^\dagger_1 - a^\dagger_2}{\sqrt{2}}, \tag{62}$$

to obtain the relatively compact one-dimensional "Moshinksky bracket"

$$\langle n_1, n_2 | n, N \rangle = \delta_{n1+n2,n+N}\frac{\sqrt{n!N!n_1!n_2!}}{2^{\frac{n_1+n_2}{2}}}\sum_i \frac{(-1)^{n-n_1+i}}{i!(N-i)!(n_1-i)!(n-n_1+i)!}, \tag{63}$$

where $n_1$, $n_2$ refer to one-dimensional oscillator wave functions associated with particles 1 and 2, and $n$, $N$ label oscillator wave functions in the corresponding relative and center-of-mass coordinates. The expectation value of the interaction then involves the integrals of Gaussians of arbitrary range against two Hermite polynomials, with additional factors of $x$, $y$, or $z$ if the $T$-violating potential in Eq. (59) is being used. The resulting expressions are straightforward.

Once the integrals have been evaluated for each Cartesian coordinate, they must be folded with the two density matrices to give the final expectation value. To do this, the code must sum over a large number of individual oscillator quantum numbers. Explicitly (and in general), the expectation value of a term in the Gaussian approximation to a finite-range potential like the ones we have been considering reads

$$E_{\text{dir}}^{\tau'k'\tau k} = \sum_{\boldsymbol{m'n'mn}} \sum_{s'u'su} \rho_{\boldsymbol{n's'},\boldsymbol{m'u'}}^{\tau'} \rho_{\boldsymbol{ns},\boldsymbol{mu}}^{\tau} G_{m'_x m_x, n'_x n_x}^{x} G_{m'_y m_y, n'_y n_y}^{y} G_{m'_z m_z, n'_z n_z}^{z} \sigma_{u's'}^{k'}(m'_y n'_y) \sigma_{us}^{k}(m_y n_y). \tag{64}$$

Here $\rho_{\boldsymbol{ns},\boldsymbol{mu}}^{\tau}$ is the density matrix for particles of type $\tau$=p,n in the basis specified by the oscillator quantum numbers $\boldsymbol{n}=(n_x n_y n_z)$ or $\boldsymbol{m}=(m_x m_y m_z)$ in the three Cartesian directions and the simplexes $s$ or $u$, $G_{m'_i m_i, n'_i n_i}^{i}$ are one-dimensional integrals representing matrix elements of a Gaussian (or a Gaussian times a coordinate) in the $i$=$x$, $y$, or $z$ directions, and $\sigma_{us}^{k}(m_y n_y)$ for $k$=0, $x$, $y$, or $z$ are the matrix elements of the Pauli matrices (with $\sigma^0 \equiv 1$) in the simplex basis (I-78) that depend on the numbers of quanta in the $y$ direction $(m_y n_y)$.

Altogether, sums in Eq. (64) require 12 independent sums over the oscillator states, i.e., about $N_0^{12}$ operations for a basis cut at $N_0$ oscillator shells, and are intractable unless performed efficiently. To reduce the number of operations, HFODD sums over the simplex quantum numbers first, resulting in the intermediate quantities

$$D_{\boldsymbol{n'm'}}^{\tau'k'} = \sum_{s'u'} \rho_{\boldsymbol{n's'},\boldsymbol{m'u'}}^{\tau'} \sigma_{u's'}^{k'}(m'_y n'_y), \tag{65}$$

$$D_{\boldsymbol{nm}}^{\tau k} = \sum_{su} \rho_{\boldsymbol{ns},\boldsymbol{mu}}^{\tau} \sigma_{us}^{k}(m_y n_y), \tag{66}$$

the computation of which requires $N_0^6$ operations. After these are stored, the sums in each Cartesian direction are done one-by-one. HFODD first computes auxiliary $y$-direction matrices

$$Y_{n_x m'_y n_z, m_x n'_y m_z}^{\tau k} = \sum_{n_y m_y} D_{n_x n_y n_z, m_x m_y m_z}^{\tau k} G_{m'_y m_y, n'_y n_y}^{y}, \tag{67}$$

a task that requires $N_0^8$ operations. Next it performs the $z$ sums, yielding

$$Z_{n_x m'_y m'_z, m_x n'_y n'_z}^{\tau k} = \sum_{n_z m_z} Y_{n_x m'_y n_z, m_x n'_y m_z}^{\tau k} G_{m'_z m_z, n'_z n_z}^{z}, \tag{68}$$

the computation of which again requires $N_0^8$ operations. The sum in the $x$ direction is postponed to save storage (the auxiliary matrices $Y_{n_x m'_y n_z, m_x n'_y m_z}^{\tau k}$ and $Z_{n_x m'_y m'_z, m_x n'_y n'_z}^{\tau k}$ are each four times larger than the original density matrix). Instead, for fixed values of the $x$ quantum numbers $n_x$, $n'_x$, $m_x$, and $m'_x$, the code first sums over the "primed" $y$ and $z$ labels, computing and storing

$$X_{n'_x n_x, m'_x m_x}^{\tau'k',\tau k} = \sum_{n'_y n'_z m'_y m'_z} D_{n'_x n'_y n'_z, m'_x m'_y m'_z}^{\tau'k'} Z_{n_x m'_y m'_z, m_x n'_y n'_z}^{\tau k}, \tag{69}$$

a procedure that once more requires $N_0^8$ operations. The final result is obtained by summing over the $N_0^4$ $x$-direction labels, giving

$$E_{\text{dir}}^{\tau'k'\tau k} = \sum_{n'_x n_x, m'_x m_x} X_{n'_x n_x, m'_x m_x}^{\tau'k',\tau k} G_{m'_x m_x, n'_x n_x}^{x}. \tag{70}$$

The separation of the Gaussian interaction into its Cartesian components is what allows the original $N_0^{12}$ terms to be reduced to $N_0^8$ terms. For comparison, the mean-field matrix elements calculated for a zero-range interaction require a sum over about $N_0^7$ terms, see I.

Figures 1 and 2 show the actual CPU times required in calculations using the Skyrme and Yukawa interactions, respectively. At $N_0 = 10$, the latter take more than two orders of magnitude longer and, moreover, their CPU times scale like $N_0^7$ rather than $N_0^4$. A local EDF clearly makes for easier computing.

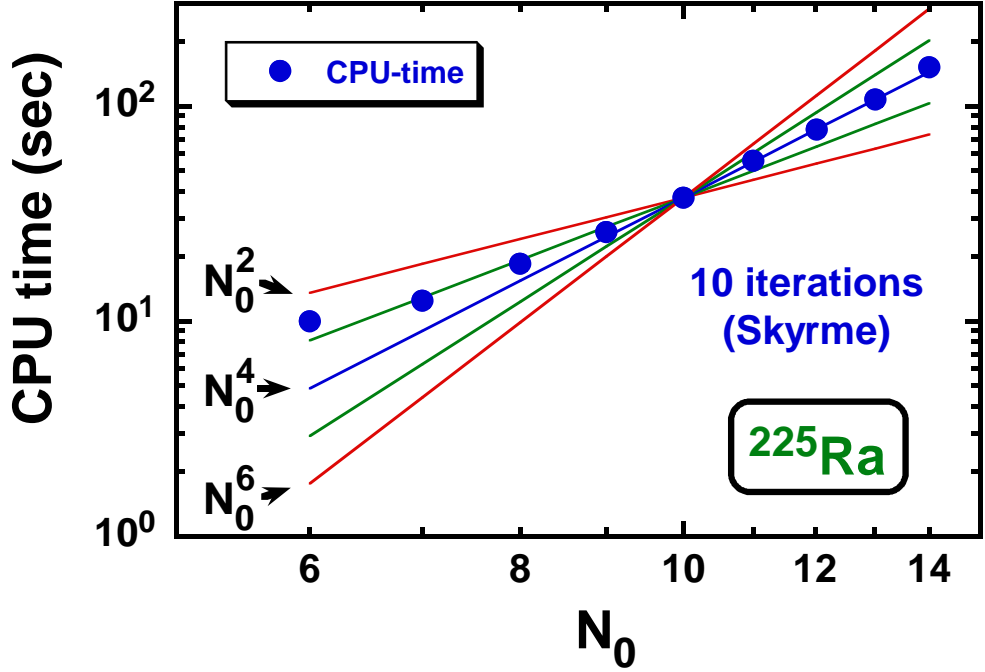

Figure 1: The HFODD CPU times required for calculations that use the standard Skyrme EDF, shown as a function of the number of HO shells $N_0$. The doubly logarithmic scale in the Figure, shows that these times scale as $N_0^4$.

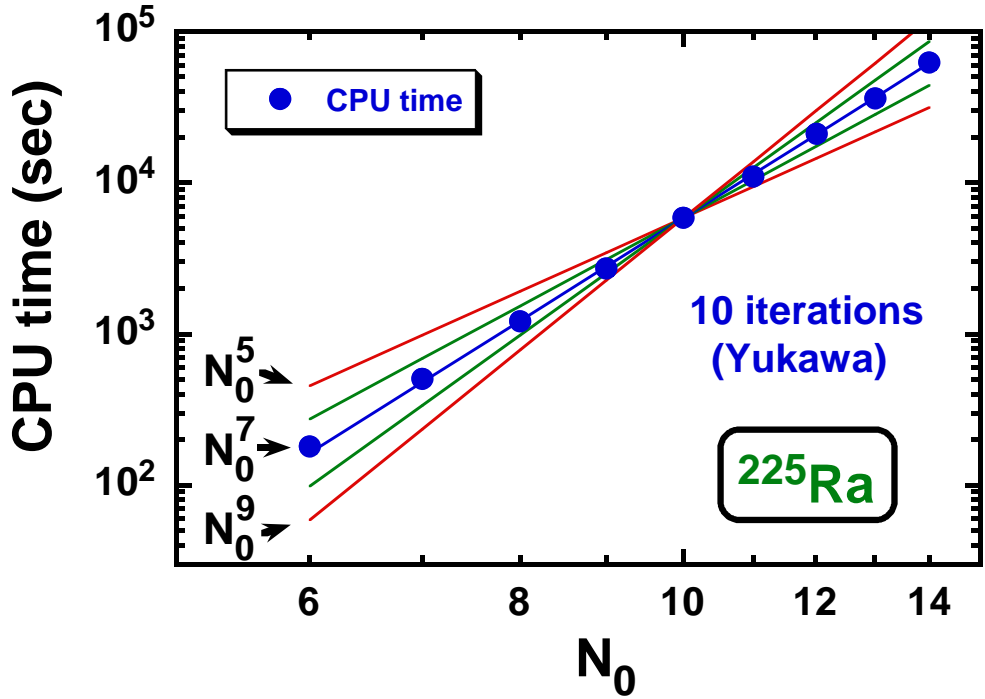

Figure 2: Same as in Figure 1 but for calculations that use the Yukawa interaction. The CPU times scale as $N_0^7$.

## 2.4 The BCS solutions for state-dependent pairing gaps

In the previous versions of the code HFODD, only the seniority pairing force has been implemented in the particle-particle channel. In version (v2.38j), we have introduced the subroutine DELPAI which solves the BCS equations for the seniority pairing force, the seniority pairing force with fixed pairing gap, as well as for the zero-range density-dependent pairing force. The latter option amounts to using the state-dependent pairing gaps.

For the seniority pairing force, the neutron and proton pairing strength constants are defined as in Ref. [21]. These values can be scaled by the multiplicative factors FACTGN and FACTGP for neutrons and protons, respectively (see Section II-3.2).

The zero-range density-dependent pairing force, Eq. (IV-14), is defined by the sets of parameters $\{V_0, V_1, \alpha\}$ or equivalently by $\{V_0, \rho_0, \alpha\}$, see Section IV-3.1. The pairing matrix elements between pairs of time-reversed states (the array GMATRI) are calculated in subroutine GINTER. After solving the BCS equations for state-dependent pairing gaps $\Delta_k$, the code calculates the average neutron and proton (spectral) gaps defined as:

$$\langle \Delta \rangle = \frac{\sum_k v_k u_k \Delta_k}{\sum_k v_k u_k} , \qquad (71)$$

were $v_k$ and $u_k$ are the BCS occupation amplitudes.

## 2.5 Calculation of Bohr deformation parameters

The shape of the nuclear surface is usually described in terms of the multipole expansion,

$$R(\theta, \phi) = R_0 \left( 1 + \sum_{\lambda=1}^{\infty} \sum_{\mu=-\lambda}^{\lambda} \alpha_{\lambda\mu} Y_{\lambda\mu}(\theta, \phi) \right) . \qquad (72)$$

In self-consistent methods, the radius $R_0$ and the deformation coefficients $\alpha_{\lambda\mu}$, $\lambda \geq 1$, have to be determined on the basis of the calculated density distribution according to some convention. The general concept is to consider a sharp-edge body of constant density $\rho_0$ and shape defined by Eq. (72), and choose $\rho_0$, $R_0$, and $\alpha_{\lambda\mu}$ so that the monopole surface moment, $Q_{00}^S$ (IV-3), and the electric multipole moments, $Q_{\lambda\mu}$ (IV-2), $\lambda \geq 0$, of the body equal those of the mean-field solution.

The moments $Q_{00}^S$ and $Q_{\lambda\mu}$ of any density distribution, $\rho(\vec{r})$, are defined as average values of the functions given by Eqs. (IV-2) and (IV-3), i.e.,

$$Q_{00}^S = \int d^3\vec{r}\, \rho\, r^2 , \qquad Q_{\lambda\mu} = a_{\lambda\mu} \int d^3\vec{r}\, \rho\, r^\lambda Y_{\lambda\mu}^* . \qquad (73)$$

The conventional factors $a_{\lambda\mu}$ are listed in Table IV-5; in particular, $a_{00} = \sqrt{4\pi}$. The monopole components, $Q_{00}^S$ and $Q_{00}$, are related to the root-mean-square radius, $R_{rms}$, $Q_{00}^S = Q_{00} R_{rms}^2$, and particle number $N$, $Q_{00} = N$. Below, the quantities pertaining to the mean-field solution are marked with bars, $\bar{R}_{rms}$, $\bar{Q}_{00}^S$ and $\bar{Q}_{\lambda\mu}$. For the reference body, Eqs. (73) take the form

$$Q_{00}^S(\rho_0, R_0, \alpha) = \frac{\rho_0}{5} \int_0^{2\pi} d\phi \int_0^\pi \sin\theta\, d\theta\, R^5(\theta, \phi) , \qquad (74)$$

$$Q_{\lambda\mu}(\rho_0, R_0, \alpha) = a_{\lambda\mu} \frac{\rho_0}{\lambda + 3} \int_0^{2\pi} d\phi \int_0^\pi \sin\theta\, d\theta\, R^{\lambda+3}(\theta, \phi)\, Y_{\lambda\mu}^*(\theta, \phi) , \qquad (75)$$

where $\alpha$ denotes the ensemble of the parameters $\alpha_{\lambda\mu}$. Thus, the equations of the problem read

$$Q^S_{00}(\rho_0, R_0, \alpha) = \bar{Q}^S_{00} , \qquad Q_{\lambda\mu}(\rho_0, R_0, \alpha) = \bar{Q}_{\lambda\mu} . \tag{76}$$

They cannot be solved analytically.

For small deformations, however, one can determine $\rho_0$ and $R_0$ from $Q^S_{00}$ and $Q_{00}$ by setting $\alpha_{\lambda\mu} = 0$ in Eqs. (74) and (75). Then, one can find $\alpha_{\lambda\mu}$ by expanding the expression in Eq. (75) up to first order in $\alpha_{\lambda\mu}$ about $\alpha_{\lambda\mu} = 0$. This leads to the linear approximation,

$$\rho_0 = \frac{3}{4\pi} \frac{\bar{Q}_{00}}{R_0^3} , \qquad R_0 = \sqrt{\frac{5}{3}} \bar{R}_{rms} , \qquad \alpha_{\lambda\mu} = \frac{4\pi}{3 a_{\lambda\mu}} \frac{\bar{Q}_{\lambda\mu}}{\bar{Q}_{00} R_0^\lambda} , \tag{77}$$

which was used in previous versions of the code HFODD, see Section III-2.9.

In version (v2.38j), exact solutions to Eqs. (76) are sought numerically, for arbitrary deformations. First, the integrals in Eqs. (74) and (75) are evaluated by using the G-L quadratures. Then, the problem is formulated in terms of fixed-point equations, which are solved by using the standard iterative method. Iterations stop when Eqs. (76) are satisfied up to an accuracy which guarantees that all the digits printed on the output are correct. It may happen for very large deformations that the procedure diverges and the exact values of the deformation parameters are not found.

Unlike in the linear approximation, the values referred to as exact depend on the value of the maximum multipolarity $\lambda_{max}$ considered in Eqs. (76). This is not a very dramatic effect, and in order to estimate the exact values for a multipolarity $\lambda$, it is recommended to perform calculations for multipolarities up to the next one of the same parity, i.e., for $\lambda_{max} = \lambda + 2$. In the code, $\lambda_{max}$ is set independently of the number of multipole moments calculated for the mean-field solution, although it must not exceed the latter one, of course.

According to the adopted convention, see Section III-2.9, the code prints the Bohr deformation parameters, $\beta_{\lambda\mu}$, which contain an additional factor of $\sqrt{2}$ introduced for $\mu \neq 0$,

$$\beta_{\lambda\mu} = \alpha_{\lambda\mu} \sqrt{2 - \delta_{\mu 0}} . \tag{78}$$

## 2.6 Constraints on the Schiff moments and scalar multipole moments

Apart from calculating and constraining the multipole moments $Q_{\lambda\mu}$ (IV-2) and surface moments $Q^S_{\lambda\mu}$ (IV-3), the code HFODD version (v2.38j) also calculates the average values of the so-called Schiff multipole moments,

$$\begin{aligned} Q^F_{\lambda\mu}(\boldsymbol{r}) &= a_{\lambda\mu} \left[ r^{\lambda+2} - \tfrac{5}{3} \langle r^2 \rangle r^\lambda \right] Y^*_{\lambda\mu}(\theta, \phi), \\ &= Q^S_{\lambda\mu}(\boldsymbol{r}) - \tfrac{5}{3} \langle r^2 \rangle Q_{\lambda\mu}(\boldsymbol{r}), \end{aligned} \tag{79}$$

where for the neutron, proton, or total Schiff moment, $\langle r^2 \rangle$ denotes the neutron, proton, or total mean-square radius, respectively. Note that the Schiff moments depend through $\langle r^2 \rangle$ on the self-consistent solution; therefore, in the given iteration of the code, values of $\langle r^2 \rangle$ are taken from the previous iteration of the code. In this way, a proper definition of the Schiff moment is ensured only after the convergence is reached. Note also that the factor of $\frac{1}{10}$, which is included in the standard definition of the Schiff moment [22, 19] is *not* included in the definition of Eq. (79).

In version (v2.38j), the code HFODD can calculate and constrain either the surface moments or the Schiff moments, according to the switch `ISCHIF`, see Section 3.5 below, but not both simultaneously.

Since the physical constraints on the Schiff moments have to be set for the proton moments only, the meaning of parameters `IFLAGS` (see keyword `SURFCONSTR` in Section IV-3.5) has been generalized in the following way: As previously, for `IFLAGS=1`, the constraints on the surface or Schiff moments pertain to the total moments, but for `IFLAGS=2` or 3 they now may pertain to the neutron or proton moments, respectively.

In version (v2.38j), the code HFODD can also set constraints on scalars $Q_\lambda$ built from the multipole moments $Q_{\lambda\mu}$:

$$Q_\lambda = a_{\lambda 0} \sqrt{\sum_{\mu=-\lambda}^{\lambda} \frac{|Q_{\lambda\mu}|^2}{a_{\lambda\mu}^2}}. \qquad (80)$$

For $\lambda = 2$, such a constraint is useful when the minimum of energy is to be found as function of the deformation $\gamma$ for a fixed value of the deformation $\beta$. Constraints on scalars are also insensitive to the overall orientation of nucleus in space.

## 2.7 Quasiparticle blocking for the HFB solutions in odd and odd-odd nuclei

In order to include pairing correlations, HFODD solves the HFB equations:

$$\mathcal{H} \begin{pmatrix} \chi & \varphi \end{pmatrix} = \begin{pmatrix} \chi & \varphi \end{pmatrix} \begin{pmatrix} E & 0 \\ 0 & -E \end{pmatrix}, \qquad (81)$$

where

$$\mathcal{H} = \begin{pmatrix} h' - \lambda & \Delta \\ -\Delta^* & -h'^* + \lambda \end{pmatrix} \qquad (82)$$

is the HFB matrix. Here, $h'$ is the s.p. Routhian operator, $\lambda$ is the Fermi energy, $\Delta$ is the antisymmetric pairing potential, and $E$ is a diagonal matrix of quasiparticle energies. In the case of even-even nuclei, $\chi$ (resp. $\varphi$) are the wavefunctions of quasiparticles with positive (resp. negative) energies. Note that these are $2M \times M$ matrices, where M is the dimension of the s.p. basis. One may write them down in terms of upper and lower components using the $M \times M$ matrices $A$ and $B$:

$$\chi = \begin{pmatrix} A \\ B \end{pmatrix}, \quad \varphi = \begin{pmatrix} B^* \\ A^* \end{pmatrix}. \qquad (83)$$

The matrices $A$ and $B$ form the matrix of the Bogolyubov transformation:

$$\mathcal{A} = \begin{pmatrix} A & B^* \\ B & A^* \end{pmatrix}. \qquad (84)$$

In the case of the so-called proper Bogolyubov transformation, for which $\det \mathcal{A} = 1$, the HFB state is a mixture of states with even number of particles only. The blocking of quasiparticle $k$ is done by replacing one column in the $\varphi$ matrix by the column which corresponds to the quasiparticle state with the opposite energy from the $\chi$ matrix. This leads to exchange of the two columns of the Bogolyubov transformation matrix $\mathcal{A}$ and $\det \mathcal{A} = -1$. Now $\mathcal{A}$ corresponds to the improper Bogulyubov transformation and the resulting HFB wave function is a superposition of states with odd number of particles only [14]. Let $A_i$ denote the i-th column of the matrix $A$. Then the corresponding HFB wave functions of blocked states have the form:

$$\chi_k = \begin{pmatrix} A_1 & \ldots & A_{k-1} & B^*_k & A_{k+1} & \ldots & A_M \\ B_1 & \ldots & B_{k-1} & A^*_k & B_{k+1} & \ldots & B_M \end{pmatrix}, \quad (85)$$

$$\varphi_k = \begin{pmatrix} B^*_1 & \ldots & B^*_{k-1} & A_k & B^*_{k+1} & \ldots & B^*_M \\ A^*_1 & \ldots & A^*_{k-1} & B_k & A^*_{k+1} & \ldots & A^*_M \end{pmatrix}. \quad (86)$$

To decide which quasiparticle should be blocked the code calculates the overlap of s.p. and quasiparticle states in the following way: overlaps of the s.p. state with the upper component and with the time-reversed (complex-conjugate) lower component of the quasiparticle state. The greater of these two is chosen. The quasiparticle state with the largest overlap with the selected s.p. state is chosen to be blocked during each iteration.

In the case of conserved simplex symmetry, the Routhian and pairing potential have the form:

$$h' = \begin{pmatrix} h'_+ & 0 \\ 0 & h'_- \end{pmatrix}, \quad \Delta = \begin{pmatrix} 0 & \Delta_+ \\ \Delta_- & 0 \end{pmatrix}. \quad (87)$$

Then, the general HFB equations decouple into two different set of independent equations. The code HFODD solves the equation:

$$\begin{pmatrix} h'_+ - \lambda & \Delta_+ \\ -\Delta^*_- & h'^*_- + \lambda \end{pmatrix} \begin{pmatrix} A_+ & B^*_+ \\ B_- & A^*_- \end{pmatrix} = \begin{pmatrix} A_+ & B^*_+ \\ B_- & A^*_- \end{pmatrix} \begin{pmatrix} E_- & 0 \\ 0 & -E_+ \end{pmatrix}, \quad (88)$$

from the solution of which, one may reconstruct the complete solution of Eq. (81):

$$\chi = \begin{pmatrix} 0 & A_+ \\ A_- & 0 \\ B_+ & 0 \\ 0 & B_- \end{pmatrix}, \quad \varphi = \begin{pmatrix} B^*_+ & 0 \\ 0 & B^*_- \\ 0 & A^*_+ \\ A^*_- & 0 \end{pmatrix}. \quad (89)$$

Blocking of the k-th quasiparticle (for $k \leq M/2$) leads to the following expressions for the wave HFB functions (85) and (86):

$$\chi_k = \left( \begin{array}{ccccccc|ccc} 0 & \ldots & 0 & B^*_{+k} & 0 & \ldots & 0 & A_{+1} & \ldots & A_{+M/2} \\ A_{-1} & \ldots & A_{-k-1} & 0 & A_{-k+1} & \ldots & A_{-M/2} & 0 & \ldots & 0 \\ B_{+1} & \ldots & B_{+k-1} & 0 & B_{+k+1} & \ldots & B_{+M/2} & 0 & \ldots & 0 \\ 0 & \ldots & 0 & A^*_{-k} & 0 & \ldots & 0 & B_{-1} & \ldots & B_{-M/2} \end{array} \right), \quad (90)$$

$$\varphi_k = \left( \begin{array}{ccccccc|ccc} B^*_{+1} & \ldots & B^*_{+k-1} & 0 & B^*_{+k+1} & \ldots & B^*_{+M/2} & 0 & \ldots & 0 \\ 0 & \ldots & 0 & A_{-k} & 0 & \ldots & 0 & B^*_{-1} & \ldots & B^*_{-M/2} \\ 0 & \ldots & 0 & B_{+k} & 0 & \ldots & 0 & A^*_{+1} & \ldots & A^*_{+M/2} \\ A^*_{+1} & \ldots & A^*_{+k-1} & 0 & A^*_{+k+1} & \ldots & A^*_{+M/2} & 0 & \ldots & 0 \end{array} \right), \quad (91)$$

where horizontal and vertical lines are introduced to distinguish $M \times M/2$ blocks. One can easily derive analogous expressions in the case of $M/2 < k \leq M$. Instead of using the blocked matrix $\varphi$ in the form of (91), the code uses instead a $2M \times (M+1)$ matrix $\varphi'$:

$$\varphi' = \left( \begin{array}{cccccc|cccc|c} B^*_{+1} & \ldots & B^*_{+k-1} & 0 & B^*_{+k+1} & \ldots & B^*_{+M/2} & 0 & \ldots & 0 & 0 \\ 0 & \ldots & 0 & 0 & 0 & \ldots & 0 & B^*_{-1} & \ldots & B^*_{-M/2} & A_{-k} \\ \hline 0 & \ldots & 0 & 0 & 0 & \ldots & 0 & A^*_{+1} & \ldots & A^*_{+M/2} & B_{+k} \\ A^*_{+1} & \ldots & A^*_{+k-1} & 0 & A^*_{+k+1} & \ldots & A^*_{+M/2} & 0 & \ldots & 0 & 0 \end{array} \right). \quad (92)$$

It can be easily seen that this matrix leads to the same generalized density matrix $\mathcal{R} = \varphi\varphi^\dagger$. In this representation, the wave function of the blocked quasiparticle is just zeroed and the wave function of the quasiparticle with opposite energy is added as an additional column.

## 2.8 The Broyden method

The HFB equations are a system of non-linear integro-differential equations, which require to be solved self-consistently: an initial guess for the density matrix and pairing density is used to generate the HF potential and pairing field and define the HFB matrix; Solving the HFB equations yield a new set of eigen-functions that are used to calculate the densities at the next step, and this loop is executed until convergence is achieved. Different criteria can be used for convergence. In HFODD, iterations are stopped when the difference between the HFB energy and the sum of s.p. energies is less than EPSITE, see I and II.

Mathematically, the self-consistent HFB equations can be viewed as a particular example of a fixed-point problem. Formally, they read:

$$\mathbf{V}^{(m)}_{\text{out}} = \mathbf{I}(\mathbf{V}^{(m)}_{\text{in}}), \quad (93)$$

where $\mathbf{V}^{(m)}_{\text{in}}$ is a vector of size $N$ containing certain initial conditions (i.e., the set of wave-functions, HF potential, matrix elements of the HFB matrix, etc.) at iteration number $m$. The solution $\mathbf{V}$ to the HFB equations satisfies: $\mathbf{V} = \mathbf{I}(\mathbf{V})$, or: $\mathbf{F}(\mathbf{V}) = \mathbf{V} - \mathbf{I}(\mathbf{V}) = 0$. By default, most of nuclear structure codes iterate the vector $\mathbf{V}$ by taking as input to iteration $m+1$ a linear combination:

$$\mathbf{V}^{(m+1)}_{\text{in}} = \alpha \mathbf{V}^{(m)}_{\text{out}} + (1-\alpha)\mathbf{V}^{(m)}_{\text{in}}. \quad (94)$$

This so-called linear mixing scheme was implemented in previous versions of HFODD.

If the function $\mathbf{F}(\mathbf{V})$ is differentiable, the roots of the equation $\mathbf{F}(\mathbf{V}) = 0$ can be found by a generalized Newton-Raphson method: this is the basis for the Broyden method to accelerate convergence. At each step $m$ of the self-consistent process, the input to the next step is computed from:

$$\mathbf{V}^{(m+1)}_{\text{in}} = \mathbf{V}^{(m)}_{\text{in}} - \mathbf{B}^{(m)}\mathbf{F}^{(m)}, \quad (95)$$

where $\mathbf{B}^{(m)}$ is an *approximation* to the inverse of the Jacobian of the vector function $\mathbf{F}(\mathbf{V})$ at iteration $m$. The exact expression of Eq. (95) was first given in [23]. The modified Broyden method is a slight modification of the original Broyden method designed to avoid storing (potentially large) $N \times N$ matrices. It was presented in [24] and applied in a variety of nuclear structure problems in [25]. In practice, it is a correction term added to Eq. (94).

The modified Broyden method was implemented in HFODD according to the formulation of [25]. In HFODD the quantities that are iterated are the components of the HF fields on the G-H integration mesh, see I. The Broyden vector therefore contains the value of the components of the self-consistent HF fields at the G-H node $(x_i, y_j, z_k)$. The size $N$ of this vector is proportional to the total number of field components and the numbers of G-H nodes along each direction NXHERM, NYHERM, NZHERM. It is given by: $N = 44 \times$ (NXHERM $\times$ NYHERM $\times$ NZHERM). For the Broyden method applied within the Lipkin-Nogami (LN) calculations, the matrix elements of the density matrices have to be stored in addition, which adds $2 \times M^2$ elements to the Broyden vector ($M$ being the size of the s.p. basis).

## 2.9 The Lipkin-Nogami method

Methods of restoring the particle-number symmetry must be implemented in studies of pairing correlations because some observables like even-odd mass staggering or pair-transfer amplitudes are influenced significantly. In addition, the quantitative impact of the particle-number restoration depends on whether the pairing correlations are strong (open-shell systems) or weak (near closed shells).

The LN regime of HFODD gives an efficient way of performing *approximate* particle-number projection (PNP) calculations. It is based on the LN method [26, 27] considered as a variant of the second-order Kamlah expansion [28, 29, 30, 31], in which the PNP energy is approximated by a simple expression,

$$\mathcal{E}_{\text{TOT}} = \mathcal{E}_{HFB} + \mathcal{E}_{\text{LN}}, \tag{96}$$

where $\mathcal{E}_{HFB}$ is the HFB energy, and $\mathcal{E}_{\text{LN}}$ is the LN correction,

$$\mathcal{E}_{\text{LN}} = -\lambda_2(\langle \hat{N}^2 \rangle - N^2) = -2\lambda_2 \text{Tr}\rho(1-\rho), \tag{97}$$

with $\lambda_2$ depending on the HFB state and representing the curvature of the energy with respect to the particle number. In the Kamlah method $\lambda_2$ is varied along with variations of the HFB state while in the LN method this variation is neglected and $\lambda_2$ is simply evaluated after each iteration in order to find the best estimate of the energy curvature.

When the HFB method is applied to a given Hamiltonian, values of $\lambda_2$ can be estimated by calculating modified mean-field potentials that are analogous to the standard HFB mean fields. However, in the spirit of the EDF approach, in HFODD is adopted an efficient algorithm [32] estimating the curvature $\lambda_2$ from the seniority-pairing expression,

$$\lambda_2 = \frac{G_{\text{eff}}}{4} \frac{\text{Tr}'(1-\rho)\kappa \; \text{Tr}'\rho\kappa - 2 \; \text{Tr}(1-\rho)^2 \rho^2}{[\text{Tr}\rho(1-\rho)]^2 - 2 \; \text{Tr}\rho^2(1-\rho)^2}, \tag{98}$$

where $\text{Tr}\mathcal{A} = \sum_n \mathcal{A}_{nn}$ stands for the trace of a matrix $\mathcal{A}$, while $\text{Tr}'\mathcal{A} = \sum_n \mathcal{A}_{n\bar{n}}$. The effective pairing strength,

$$G_{\text{eff}} = -\frac{\bar{\Delta}^2}{E_{\text{pair}}}, \tag{99}$$

is determined from the HFB pairing energy,

$$E_{\text{pair}} = -\frac{1}{2}\text{Tr}\Delta\kappa^*, \tag{100}$$

and the average pairing gap,

$$\bar{\Delta} = \frac{\text{Tr}'\Delta\rho}{\text{Tr}\rho}. \tag{101}$$

All quantities defining energy $\mathcal{E}_{\text{LN}}$, Eq. (97), and $\lambda_2$, Eqs. (98)-(101), depend on the self-consistent solution, and the microscopic interaction. In HFODD, they are calculated in the subroutine LIPCOR using the canonical representation for the density matrices in terms of the canonical occupation amplitudes $U_k$ and $V_k$.

As a result of the LN calculations, pairing correlations never collapse, which is also the case of the exact PNP before variation. A comparison of different HFB particle-number projection results is further discussed in Ref. [32].

## 2.10 Exact Coulomb exchange

In evaluating the exact Coulomb exchange mean fields and energies we use the methods developed for the Gogny interaction, as described in Appendix A.5 of Ref. [33]. They are based on expanding the Coulomb interaction into a sum of Gaussians, which allows using the same infrastructure as developed for the Yukawa interaction discussed in Section 2.3. In particular, we use the identities:

$$\frac{1}{r} = \frac{2}{\sqrt{\pi}} \int_0^\infty d\alpha \exp(-\alpha^2 r^2) = \frac{2}{L\sqrt{\pi}} \int_0^1 d\xi \left(1 - \xi^2\right)^{-3/2} \exp\left(-\frac{\xi^2 r^2}{L^2 (1 - \xi^2)}\right), \tag{102}$$

where the second integral has been reduced to a finite domain by the substitution $\alpha = \frac{\xi}{L}(1-\xi^2)^{-1/2}$ and $L$ stands for the largest of the three HO lengths $L_\mu = \sqrt{\hbar/m\omega_\mu}$, $\mu = x, y, z$. By using the G-L quadratures we can now present the Coulomb interaction as a finite sum of $N_C$ Gaussians:

$$\frac{1}{r} = \sum_{i=1}^{N_C} A_i \exp\left(-a_i r^2\right), \tag{103}$$

where constants $A_i$ and $a_i$ depend on the G-L weights $W_i$ and nodes $\xi_i$ as

$$A_i = \frac{2W_i}{L\sqrt{\pi}} \left(1 - \xi_i^2\right)^{-3/2} \quad, \quad a_i = \frac{\xi_i^2}{L^2 (1 - \xi_i^2)}. \tag{104}$$

It turns out that this method provides for very precise and rapidly converging results for the exact Coulomb energies. This is illustrated in Fig. 3, where the error in the exact Coulomb exchange energy, $\Delta E_{\text{exc}}$, is plotted as function of the number of Gaussians $N_C$. A quite precise estimate is obtained for $N_C = 7$ Gaussians and the machine precision is obtained by doubling this number (along with doubling the CPU time). For $N_C = 7$, these CPU times are shown in Fig. 4, which illustrates the fact that these times scale with the number of HO shells as $N_0^7$.

## 2.11 Utility options

The following utility options have been implemented:

1. In all the keywords, the minus character "-" can always be used in place of the underscore character "_" and *vice versa*.

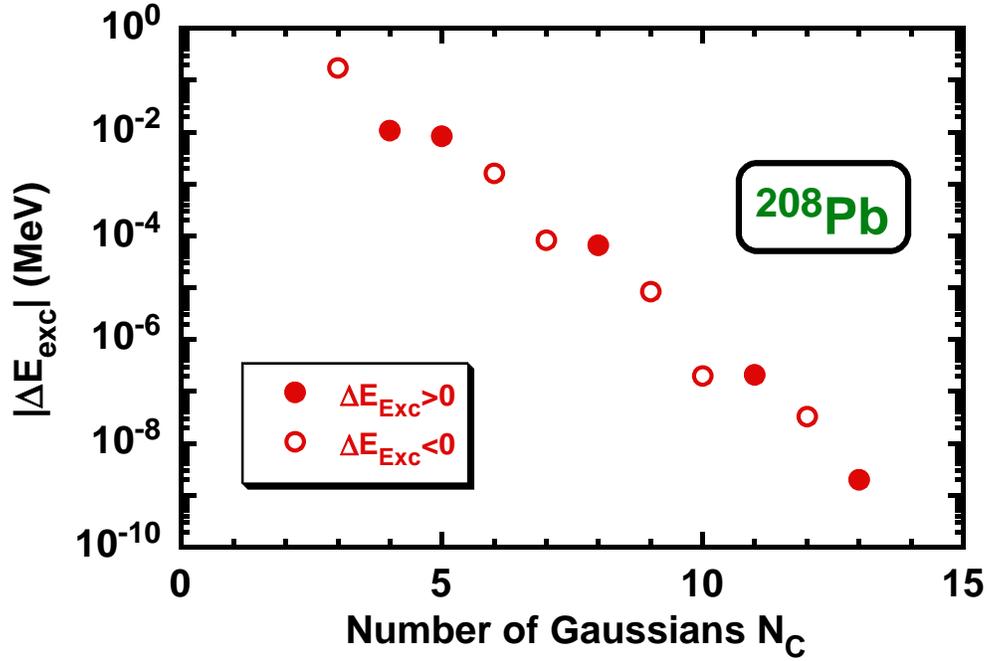

Figure 3: Errors in the exact Coulomb exchange energy, $\Delta E_{\text{exc}}$, plotted in function of the number of Gaussians $N_C$ used in Eq. (103).

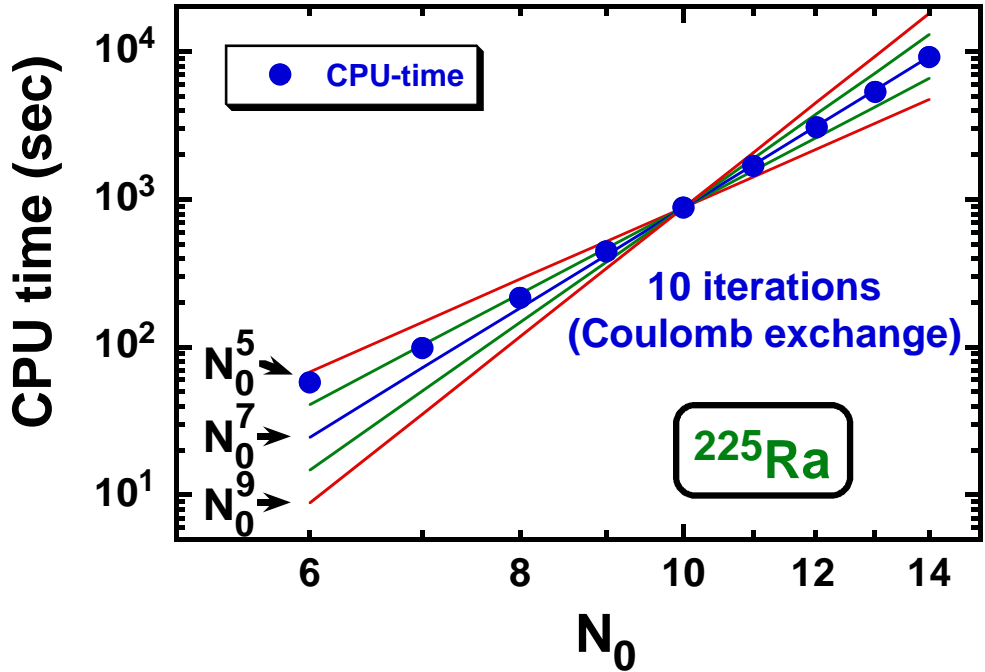

Figure 4: Same as in Figure 1 but for the calculations that determine the exact Coulomb exchange energies. The CPU times scale as $N_0^7$.

2. An external HO potential can be added to the self-consistent mean field to calculate properties of an atomic gas in a trap, see the keyword INSERT_HO below.

3. Arbitrary values can be added to the coupling constants, see the keywords EVE_ADD_TS, ODD_ADD_TS, EVE_ADD_PM, and ODD_ADD_PM below.

4. Calculations with fixed values of the Fermi energies can be performed, see the keywords FIXFERMI_N and FIXFERMI_P below.

5. For the HFB calculations, eigenvalues of the HFB mean-field s.p. Hamiltonian or Routhian can be printed, see the keyword HFBMEANFLD below.

6. Within the HF calculations, the filling approximation can be used, see the keywords FILSIG_NEU and FILSIG_PRO below.

7. Constraints on the intrinsic spin only can be used, see the keyword NORBCONSTR below.

8. Various one-line data can be printed during the iteration, see the keyword ONE_LINE below.

9. Integrals of several symmetry-violating terms can be printed, see the keyword PRINT_VIOL below.

10. The Nilsson labels with respect to the $x$, $y$, or $z$ axes can be printed, see the keyword NILSSONLAB below.

11. Matrix elements of the mean-field Hamiltonian can be saved on the disk, see the keywords FIELD_SAVE, FIELD_OLD, REC_FIELDS, and CONTFIELDS below.

### 2.12 Corrected errors

*2.12.1 Euler angles.* The Euler angles have been calculated for the quadrupole moment, i.e., for the complex conjugate spherical harmonics, instead of the complex conjugate quadrupole moment, i.e., for the spherical harmonics themselves.

*2.12.2 Predefinition of* STIFFA. In subroutine PREDEF, the value of variable STIFFA has been incorrectly predefined to 0.01 instead of 0.00.

## 3 Input data file

The structure of the input data file has been described in Section II-3; in the present version (v2.38j) of the code HFODD this structure is exactly the same. All previous items of the input data file remain valid, and several new items are added, as described in Secs. 3.1–3.8.

Together with the FORTRAN source code in the file hfodd.f, several examples of the input data files are provided. File dy152-f.dat contains all the valid input items, and the input data are identical to the default values. Therefore, the results of running the code with the input data file dy152-f.dat are identical to those obtained for the input data file containing only one line with the keyword EXECUTE.

File `ge064-a.dat` contains the input data that allow for determining the rotating triaxial state in the nucleus $^{64}$Ge and then performing a 3D AMP of this state. This data file comprises four runs of the code: (i) starting from the Nilsson initial potential, the code performs 20 iterations with constrained values of quadrupole moments $Q_{20}$ and $Q_{22}$, such that the initial deformation and orientation of the nucleus is determined, (ii) after releasing the constraints, the triaxial minimum is converged, (iii) the rotating solution for $I = 6$ is converged by setting the angular frequency of $\hbar\omega = 0.575$ MeV, and (iv) the AMP of the rotating solution is performed. Small numbers of G-T (10) and G-L nodes (10), see Section 2.1, which are used in this run, do not allow for a precise AMP resolution of high angular momenta. This data file is only meant to provide an example of rapid calculation.

File `ge064-b.dat` contains the input data that allow for a correct AMP, with higher numbers of G-T (40) and G-L nodes (40), but its execution requires a CPU time which is $4^3$=64 times longer (about 200h). Alternatively, one can run in parallel 40 jobs by executing the input data files `ge064-c.dat` with characters NUASTA replaced by integers from 1 to 40.

File `sn120-b.dat` contains the input data that allow for determining the ground state in the nucleus $^{120}$Sn with the Lipkin-Nogami corrections taken into account.

File `ge064-a.dat` is reproduced in section TEST RUN INPUT below. Files `ge064-a.out` and `sn120-b.out` contain results of executing code HFODD version (v2.38j) for the two corresponding input files. Selected lines from the output file `ge064-a.out` are reproduced in section TEST RUN OUTPUT below.

## 3.1 Interaction

**Keyword:** COULOMBPAR

  7, 1, 1 = ICOTYP, ICOUDI, ICOUEX

For ICOUDI=0, 1, or 2, the Coulomb direct energy and Coulomb mean field are neglected, calculated by using the Green-function method, see Section I-5, or calculated by using the Gaussian-expansion method, see Section 2.10, respectively. Similarly, for ICOUEX=0, 1, or 2, the Coulomb exchange energy and Coulomb mean field are neglected, calculated by using the Slater approximation (I-19) or calculated exactly by using the Gaussian-expansion method, see Section 2.10, respectively. For the Gaussian-expansion method, that is, for ICOUDI=2 or ICOUEX=2, positive values of ICOTYP denote the numbers of G-L nodes used in the integral of Eq. (103). For ICOUDI=2 or ICOUEX=2, the iteration can later be smoothly continued (IFCONT=1, see the keyword CONTFIELDS) only by saving the matrix elements of the mean field, that is, by requesting IWRIFI=1, see the keyword FIELD_SAVE. Therefore, ICOUDI=2 or ICOUEX=2 and ICONTI=1 requires IFCONT=1.

**Keyword:** PAIR_MATRI

  1, 0, 0, 0 = IDESTA, IDEMID, IDESTO, IDEDIS

For IDESTA=1, the pairing matrix elements, required for the BCS pairing calculations with state-dependent pairing gaps (IPABCS=3), are calculated in the first iteration. At present, only the value of IDESTA=1 is allowed, because the option of storing the pairing matrix elements is not yet implemented. For IDEMID=1, the pairing matrix elements are calculated in the middle iterations, for IDESTO=1 in the last iteration, and/or for IDEDIS=$n$, in every $n$-th iteration.

**Keyword:** INSERT_HO

$$0 = \text{IPOTHO}$$

For `IPOTHO=1`, an external HO potential is added to the self-consistent mean field. Parameters of the potential are identical to those defining the HO basis.

**Keyword:** `EVE_ADD_TS`

      0., 0., 0., 0., 0., 0., 0., 0., 0., 0., 0., 0.
      ARHO_T, ARHO_S, ARHODT, ARHODS, ALPR_T, ALPR_S,
              ATAU_T, ATAU_S,
              ASCU_T, ASCU_S,
              ADIV_T, ADIV_S

By using this item, the coupling constants corresponding to a given Skyrme parameter set can be shifted by arbitrary values. The time-even coupling constants in the total-sum representation (I-14) are modified by adding the 12 numbers from `ARHO_T` to `ADIV_S`. By setting the scaling factors `SRHO_T` to `SDIV_S` equal to zero, see the keyword `EVE_SCA_TS`, one can input here a new set of the coupling constants. The name convention of variables is here the same as for many other variables in the code HFODD, see the keyword `EVE_SCA_TS`.

**Keyword:** `ODD_ADD_TS`

      0., 0., 0., 0., 0., 0., 0., 0., 0., 0., 0., 0.
      ASPI_T, ASPI_S, ASPIDT, ASPIDS, ALPS_T, ALPS_S,
              ACUR_T, ACUR_S,
              AKIS_T, AKIS_S,
              AROT_T, AROT_S

Same as above but for the time-odd coupling constants.

**Keyword:** `EVE_ADD_PM`

      0., 0., 0., 0., 0., 0., 0., 0., 0., 0., 0., 0.
      ARHO_P, ARHO_M, ARHODP, ARHODM, ALPR_P, ALPR_M,
              ATAU_P, ATAU_M,
              ASCU_P, ASCU_M,
              ADIV_P, ADIV_M

Same as above but for the time-even coupling constants in the isoscalar-isovector representation (I-15). The total-sum additive factors are used first, and the isoscalar-isovector additive factors are used afterwards.

**Keyword:** `ODD_ADD_PM`

      0., 0., 0., 0., 0., 0., 0., 0., 0., 0., 0., 0.
      ASPI_P, ASPI_M, ASPIDP, ASPIDM, ALPS_P, ALPS_M,
              ACUR_P, ACUR_M,
              AKIS_P, AKIS_M,
              AROT_P, AROT_M

Same as above but for the time-odd coupling constants in the isoscalar-isovector representation.

**Keyword:** `YUKAWATERM`

      0.7045, 4.7565, 1.0, 0.0, 0.0, 0.0, 1, 0
      PIMASS, PNMASS, YUKAGT, YUKAG0, YUKAG1, YUKAG2, IYUTYP, I_YUKA

For I_YUKA>0, the code calculates the average values of the time-reversal- and parity-violating Yukawa interaction (59), with the pion mass ($m_\pi$) of PIMASS and the nucleon mass ($m_N$) of PNMASS. If values of zero are read, variables PIMASS and PNMASS remain unchanged. Variables YUKAGT, YUKAG0, YUKAG1, and YUKAG2 correspond, respectively, to the coupling constants $g$, $\bar{g}_0$, $\bar{g}_1$, and $\bar{g}_2$. For I_YUKA=2 or 3, the direct matrix elements of the Yukawa interaction (59) are, in addition, added to the self-consistent mean field. For I_YUKA=2 or 4, the exchange matrix are added. For IYUTYP=1, expression (61) is used, while for IYUTYP=2, an analogous six-Gaussian expression is used without the short-range correction (60), that is, for $f(r) = 1$. For I_YUKA=0, all these input data are ignored and the Yukawa interaction is not taken into account.

## 3.2 Symmetries

**Keyword:** FILSIG_NEU

    2, 2, 2, 2,   1, 1, 1, 1,   0, 0, 0, 0 =
    KPFILG(0,0,0), KPFILG(0,1,0), KPFILG(1,0,0), KPFILG(1,1,0),
    KHFILG(0,0,0), KHFILG(0,1,0), KHFILG(1,0,0), KHFILG(1,1,0),
    KOFILG(0,0,0), KOFILG(0,1,0), KOFILG(1,0,0), KOFILG(1,1,0)

These parameters govern the filling approximation for the neutron s.p. parity–signature configurations. Matrices KPFILG contain the indices of particle states in the four parity–signature blocks denoted by (+,+), (+,−), (−,+), and (−,−), of given (parity,signature) combinations, i.e., $(\pi,r) = (+1,+i)$, $(+1,-i)$, $(-1,+i)$, and $(-1,-i)$, respectively. Matrices KHFILG contain analogous indices of hole states, and matrices KOFILG contain numbers of particles put into the states between KHFILG and KPFILG, by using for them partial occupation factors of KOFILG/(KPFILG−KHFILG+1)/2. For KOFILG=0, in the given parity–signature block the filling approximation is inactive. The filling approximation is incompatible with pairing correlations, IPAIRI=1.

**Keyword:** FILSIG_PRO

    2, 2, 2, 2,   1, 1, 1, 1,   0, 0, 0, 0 =
    KPFILG(0,0,1), KPFILG(0,1,1), KPFILG(1,0,1), KPFILG(1,1,1),
    KHFILG(0,0,1), KHFILG(0,1,1), KHFILG(1,0,1), KHFILG(1,1,1),
    KOFILG(0,0,1), KOFILG(0,1,1), KOFILG(1,0,1), KOFILG(1,1,1)

Same as above but for the proton s.p. parity–signature configurations.

**Keyword:** BCS

    −1 = IPABCS

Parameter IPABCS defines the type of BCS paring calculations. For IPABCS=0, 1, 2, or 3, the BCS pairing calculations are, respectively, not performed, performed with the seniority pairing force, performed with fixed pairing gaps, or performed with the state-dependent pairing gaps. Value of IPABCS=−1 is allowed for the sake of compatibility with earlier versions of the code, before (v2.13f). Then, IPABCS=0 is set for IPAHFB=1 and IPABCS=IPAIRI is set for IPAHFB=0.

For IPABCS=1 the code HFODD solves the BCS equations with the neutron and proton pairing strengths defined in Ref. [21]. These values can be modified by defining the multiplicative factors `FACTGN` and `FACTGP` for neutrons and protons, respectively. For `IPABCS=2`, the BCS pairing calculations are performed with fixed values of the neutron or proton pairing gaps equal to `DELFIN` or `DELFIP`, respectively. For `IPABCS=3`, the BCS pairing calculations are performed with state-dependent pairing gaps corresponding to the pairing matrix elements (see the keyword `PAIR_MATRI`) calculated for the contact forces defined in the same way as for the HFB pairing calculations.

Positive values of `IPABCS` require `IPAIRI=1` and `IPAHFB=0` and are incompatible with rotations `IROTAT=1` or broken simplex `ISIMPY=0`. `IPABCS=2` requires `IDEFIN=IDEFIP=1`.

**Keyword:** `INI_INVERS`

$$0, 0 = \text{INIINV}, \text{INIKAR}$$

Allowed values of `INIKAR=0, 1, 2,` or 3 and `INIINV=0, 1, 2,` or 3 correspond to the following $D_{2h}^T$ transformations:

|           | INIKAR=0  | INIKAR=1    | INIKAR=2    | INIKAR=3    |
|-----------|-----------|-------------|-------------|-------------|
| INIINV=0  | $\hat{I}$ | $\hat{R}_x$ | $\hat{R}_y$ | $\hat{R}_z$ |
| INIINV=1  | $\hat{P}$ | $\hat{S}_x$ | $\hat{S}_y$ | $\hat{S}_z$ |
| INIINV=2  | $\hat{T}$ | $\hat{R}_x^T$ | $\hat{R}_y^T$ | $\hat{R}_z^T$ |
| INIINV=3  | $\hat{P}^T$ | $\hat{S}_x^T$ | $\hat{S}_y^T$ | $\hat{S}_z^T$ |

where $\hat{P}$ is the space inversion, $\hat{T}$ is the time reversal, $\hat{R}_k$ is the signature (rotation by angle $\pi$ about the $k = x, y,$ or $z$ axis), and $\hat{P}^T = \hat{P}\hat{T}$, $\hat{S}_k = \hat{P}\hat{R}_k$ (simplex), $\hat{R}_k^T = \hat{T}\hat{R}_k$ ($k$-signature$^T$), and $\hat{S}_k^T = \hat{T}\hat{S}_k$ ($k$-simplex$^T$). For `INIKAR=INIINV=0`, no transformation is performed and this option is inactive.

Transformations are performed at the level of the densities, after the first iteration. As a security measure, nonzero values of `INIKAR` and `INIINV` require `NOITER=1`. Such values also require `SLOWEV=SLOWOD=SLOWPA=0.0`, so as not to mix the old and new potentials corresponding to the original and transformed densities, respectively. They are also incompatible with `IPRGCM`$\neq$0.

A given $D_{2h}^T$ transformation must be accompanied by the correspondingly broken symmetries, that is, `INIINV=1` or 3 requires `IROTAT=1` and `INIINV=2` or 3 requires `IPARTY=0`.

**Keyword:** `INI_ROTAT`

$$0.0, 0.0, 0.0, 0 = \text{ALPINI}, \text{BETINI}, \text{GAMINI}, \text{INIROT}$$

For `INIROT=1`, the wave functions are rotated by the Euler angles $\alpha$, $\beta$, and $\gamma$ corresponding, respectively, to `ALPINI`, `BETINI`, and `GAMINI` (all in degrees). Transformations are performed at the level of the densities, after the first iteration. As a security measure, `INIROT=1` requires `NOITER=1`. It also requires `SLOWEV=SLOWOD=SLOWPA=0.0`, so as not to mix the old and new potentials corresponding to the original and transformed densities, respectively. `INIROT=1` is incompatible with `IPRGCM`$\neq$0 and requires a spherical HO basis of $\hbar\omega_x = \hbar\omega_y = \hbar\omega_z$.

A rotation about the $z$ axis must be accompanied by the broken signature, that is, `ALPINI`$\neq$0 or `GAMINI`$\neq$0 requires `ISIGNY=0`.

**Keyword:** PROJECTGCM

$$\begin{array}{ccc} 0,\,0,\,0, & 1,\,1,\,0, & 1,\,1,\,0 \\ \text{IPRGCM, IPROMI, IPROMA,} & \text{NUAKNO, NUBKNO, KPROJE,} \\ & \text{IFRWAV, ITOWAV, IWRWAV} \end{array}$$

For IPRGCM=1 and 2, the code calculates the diagonal and non-diagonal GCM kernels, respectively. In addition, for NUAKNO≠1 or NUBKNO≠1, the AMP kernels are calculated, as described in Section 2.1, although at present this option is not yet available for non-diagonal kernels of IPRGCM=2. The AMP is performed for doubled angular momenta from IPROMI to IPROMA and requires a spherical HO basis of $\hbar\omega_x = \hbar\omega_y = \hbar\omega_z$. For even (odd) particle number IN_FIX+IZ_FIX, IPROMI, IPROMA, and KPROJE must be even (odd). For IPRGCM=1 and 2, the present version (v2.38j) also requires IROTAT=1 and IPAIRI=0.

The AMP has been tested up to the values of angular momenta of $70\hbar$, and therefore, IPROMA must not be larger than 2*70=140. After further tests, higher values could be used by increasing the parameter JMAX=70 in function DSMALG, which calculates the Wigner functions, see Eqs. (1), (4), and (5).

NUAKNO is the number of G-T nodes, which are used to perform integrations over the $\alpha$ and $\gamma$ Euler angles. For NUAKNO=1, these integrations are not performed (1D AMP) and the states are assumed to be axial with the doubled projection of the angular momentum on the $z$ axis equal to KPROJE. NUBKNO is the number of G-L nodes, which are used to perform integrations over the $\beta$ Euler angle. For NUAKNO>1 and NUBKNO>1, a full 3D AMP is performed and the value of KPROJE is ignored. NUAKNO>1 requires ISIMPY=0 and ISIGNY=0. IPROMA must be larger than the absolute value of KPROJE.

For IPRGCM=2, the code calculates the GCM kernels between the states labeled by three-digit indices from "000" to "999". Indices of the "left" states vary between IFRWAV and ITOWAV, and these states are read from the disc. The index of the "right" state equals to ITOWAV, and this state is equal to the current state. In addition, for IWRWAV=1, the current state is saved on the disc with the index of ITOWAV. This allows for a simultaneous calculation of the "right" state along with calculating its kernels with all previously calculated and stored "left" states. The states can also be stored on disc without calculating kernels in the given run, that is, by setting the value of ITOWAV along with IWRWAV=1 and IPRGCM=0. Names of files on the disc are composed by concatenating the three-digit index, "-", and FILWAV.

**Keyword:** SAVEKERNEL

$$0 = \text{ISAKER}$$

For ISAKER=1, the code attempts reading the kernel files Nxxx-Lyyy-Rzzz-//FILKER, where the three-digit indices are:

- xxx is the consecutive index of the kernel file,
- yyy is the index of the left wave function,
- zzz is the index of the right wave function.

In the work directory, the file names for all indices xxx are scanned, starting from 001. The kernels stored in these files are read into memory and are not recalculated. The kernels that have not been found in the kernel files are calculated and stored in the kernel file with the lowest available index xxx. In this way, one can submit many parallel jobs, see the keyword PARAKERNEL, that calculate kernels for different values of the Euler angles $\alpha$, $\beta$, and $\gamma$. The results are then collected in different kernel files with indices xxx attributed automatically. If any of the jobs is terminated before completing its task, the same input data can be resubmitted and the calculation automatically continues from the point where it has been interrupted. Once all the kernels will have been calculated, which requires a large CPU time, the AMP can be performed within a very small CPU time by reading, again automatically, all the created kernel files. ISAKER=1 requires IPRGCM>0

**Keyword:** CHECKERNEL

$$1 = \text{ICHKER}$$

The names of kernel files are saved within these files. As a security measure, when reading the kernel files, their names are cross-checked against the saved information. This cross-checking can be deactivated by using ICHKER=0. This option is useful whenever the kernel files have been renamed for any reason.

**Keyword:** PARAKERNEL

$$0, 1, 1, 1, 1 = \text{IPAKER, NUASTA, NUASTO, NUGSTA, NUGSTO}$$

For IPAKER=1, the code only calculates kernels for different values of the Euler angles $\alpha$, $\beta$, and $\gamma$ and the AMP is suspended. Calculations are performed for the G-T nodes in the Euler angle $\alpha$ from NUASTA to NUASTO, for those in the Euler angle $\gamma$ from NUGSTA to NUGSTO, and for all the G-L nodes in the Euler angle $\beta$, that is, from 1 to NUBKNO. For IPAKER=1, to save memory the code can be compiled with IPARAL=1. IPAKER=1 requires IPRGCM>0 and ISAKER=1. Values of NUASTA, NUASTO, NUGSTA, NUGSTO must all be between 1 and NUAKNO and must be ordered as NUASTA$\leq$NUASTO and NUGSTA$\leq$NUGSTO.

**Keyword:** TRANSITION

$$2, 1, 0 = \text{NMURED, NMARED, NSIRED}$$

Maximum numbers of transition electric, magnetic, and surface or Schiff moments, respectively, for which proton kernels and reduced matrix elements are calculated, printed, and stored in the kernel files. NMARED and NSIRED must not exceed NMURED. For NMURED=0, NMARED=0, or NSIRED, the corresponding proton kernels and reduced matrix elements are not calculated.

**Keyword:** CUTOVERLAP

$$0, 10^{-10}, 1. = \text{ICUTOV, CUTOVE, CUTOVF}$$

For ICUTOV=0, parameters CUTOVE and CUTOVF are ignored and the collective states for the $K$-mixing calculation, Eq. (10), are selected by their norm eigenvalues $n_m$ being larger than the negative of the smallest norm eigenvalue OVEMIN. For ICUTOV=1, the collective states are selected by their norm eigenvalues being larger than CUTOVE+CUTOVF*OVEMIN.

**Keyword:** LIPKIN

$$0, 0 = \text{LIPKIN, LIPKIP}$$

For LIPKIN=1 and/or LIPKIP=1, the Lipkin-Nogami corrections are included for neutrons and/or protons, respectively, see Section 2.9. At present, LIPKIN=1 or LIPKIP=1 requires IPAHFB=1.

**Keyword:** `INI_LIPKIN`
$$0.1, 0.1 = \text{FE2INI(0)}, \text{FE2INI(1)}$$
For `ICONTI=0` or `ICONTI=1` and `ILCONT=0`, the Lipkin-Nogami calculations are started by using the initial values of the neutron and proton Lipkin-Nogami parameters $\lambda_2$ (98), `FE2INI(0)` and `FE2INI(1)`, respectively. For `ICONTI=0` or `ILCONT=0`, values of `FE2INI` are ignored.

**Keyword:** `FIXLAMB2_N`
$$0.1, 0 = \text{FE2FIN}, \text{IF2FIN}$$
For `IF2FIN=1`, the neutron Lipkin-Nogami calculations are performed by using the fixed value of the neutron Lipkin-Nogami parameter $\lambda_2$ (98) equal to `FE2FIN`. `IF2FIN=1` requires `LIPKIN=1`.

**Keyword:** `FIXLAMB2_P`
$$0.1, 0 = \text{FE2FIP}, \text{IF2FIP}$$
Same as above but for the proton Lipkin-Nogami calculations.

**Keyword:** `SLOWLIPKIN`
$$0.5 = \text{SLOWLI}$$
`SLOWLI` gives the value of the mixing fraction $\epsilon$ used for the Lipkin-Nogami parameter $\lambda_2$ (98), in analogy to the `SLOWEV`, `SLOWOD` and `SLOWPA` parameters.

**Keyword:** `FIXFERMI_N`
$$-8.0, 0 = \text{FERFIN}, \text{IFEFIN}$$
For `IFEFIN=1`, the HFB pairing calculations are performed with a fixed value of the neutron Fermi energy equal to `FERFIN`. At present, `IFEFIN=1` requires `IPAHFB=1`.

**Keyword:** `FIXFERMI_P`
$$-8.0, 0 = \text{FERFIP}, \text{IFEFIP}$$
Same as above but for the proton HFB pairing calculations.

## 3.3 Configurations

**Keyword:** `BLOCKSIZ_N`
$$1, 0 = \text{INSIZN}, \text{IDSIZN}$$
For $|\text{IDSIZN}|=1$, the code performs the neutron quasiparticle blocking calculations in the case when no symmetries are conserved, see Section 2.7. For `IDSIZN=+1` or $-1$, the blocked quasi-particle state is selected by having the largest overlap with the `INSIZN`-th neutron s.p. eigenstate of the HFB mean-field Routhian or with its time-reversed partner, respectively. Note that for rotating states, the time-reversed eigenstate is not necessarily an eigenstate of the Routhian. $|\text{IDSIZN}|=1$ requires `ISIMPY=0`, `IPARTY=0`, `IPAHFB=1`, and `IROTAT=1`.

**Keyword:** `BLOCKSIZ_P`
$$1, 0 = \text{INSIZP}, \text{IDSIZP}$$
Same as above but for the proton quasiparticle blocking. For odd-odd nuclei, neutron and proton quasiparticles can be simultaneously blocked.

**Keyword:** `BLOCKSIM_N`
$$1, 0, 0 = \text{INSIMN}, \text{IRSIMN}, \text{IDSIMN}$$
For $|\text{IDSIMN}|=1$, the code performs the neutron quasiparticle blocking calculations in the case when simplex is conserved, see Section 2.7. For `IDSIMN=+1` or $-1$, the blocked quasiparti-cle state is selected by having the largest overlap with the INSIMN-th neutron s.p. eigenstate of the HFB mean-field Routhian in a given simplex block or with its time-reversed partner, respectively. The simplex of the state, $+i$ or $-i$, is defined by IRSIMN=0 or 1, respectively. |IDSIMN|=1 requires ISIMPY=1, IPARTY=0, IPAHFB=1, and IROTAT=1.

**Keyword:** BLOCKSIM_P

$$1, 0, 0 = \text{INSIMP, IRSIMP, IDSIMP}$$

Same as above but for the proton quasiparticle blocking. For odd-odd nuclei, neutron and proton quasiparticles can be simultaneously blocked.

**Keyword:** BLOCKSIQ_N

$$1, 0, 0 = \text{INSIQN, IPSIQN, IDSIQN}$$

For |IDSIQN|=1, the code performs the neutron quasiparticle blocking calculations in the case when parity is conserved. For IDSIQN=+1 or $-1$, the blocked quasiparticle state is selected by having the largest overlap with the INSIQN-th neutron s.p. eigenstate of the HFB mean-field Routhian in a given parity block or with its time-reversed partner, respectively. The parity of the state, $+1$ or $-1$, is defined by IPSIQN=0 or 1, respectively. |IDSIQN|=1 requires ISIMPY=0, IPARTY=1, IPAHFB=1, and IROTAT=1.

**Keyword:** BLOCKSIQ_P

$$1, 0, 0 = \text{INSIQP, IPSIQP, IDSIQP}$$

Same as above but for the proton quasiparticle blocking. For odd-odd nuclei, neutron and proton quasiparticles can be simultaneously blocked.

**Keyword:** BLOCKSIG_N

$$1, 0, 0, 0 = \text{INSIGN, IPSIGN, ISSIGN, IDSIGN}$$

For |IDSIGN|=1, the code performs the neutron quasiparticle blocking calculations in the case when parity and signature are conserved. For IDSIGN=+1 or $-1$, the blocked quasiparticle state is selected by having the largest overlap with the INSIGN-th neutron s.p. eigenstate of the HFB mean-field Routhian in a given parity–signature block or with its time-reversed partner, respectively. The parity of the state, $+1$ or $-1$, is defined by IPSIGN=0 or 1, respectively. The signature of the state, $+i$ or $-i$, is defined by ISSIGN=0 or 1, respectively. |IDSIGN|=1 requires ISIMPY=1, IPARTY=1, IPAHFB=1, and IROTAT=1.

**Keyword:** BLOCKSIG_P

$$1, 0, 0, 0 = \text{INSIGP, IPSIGP, ISSIGP, IDSIGP}$$

Same as above but for the proton quasiparticle blocking. For odd-odd nuclei, neutron and proton quasiparticles can be simultaneously blocked.

**Keyword:** BLOCKFIX_N

$$0, 0 = \text{IFIBLN, INIBLN}$$

For IFIBLN=1, the neutron quasiparticle blocking is based on calculating overlaps with a fixed s.p. wave function. The method is based on beginning the iteration by defining the number of the s.p. state, that is, INSIGN, INSIMN, INSIQN, or INSIZN, depending on the selected symmetry conditions. Then, for INIBLN=1 this one s.p. wave function is stored in memory and on the record file, and in consecutive iterations the overlaps are calculated with respect to this fixed s.p. wave function. Therefore, subsequent changes in the ordering and structure of s.p. states do not affect the blocking mechanism. For INIBLN=0, in the first iteration this fixed s.p. wave function is not initialized, but it is read from the record file. IFIBLN=1 requires |IDSIGN|=1, |IDSIMN|=1,

|IDSIQN|=1, or |IDSIZN|=1, depending on the selected symmetry conditions. IFIBLN=1 and INIBLN=0 requires ICONTI=1.

**Keyword:** BLOCKFIX_P
$$0, 0 = \text{IFIBLP, INIBLP}$$
Same as above but for the proton quasiparticle blocking.

## 3.4 Miscellaneous parameters

**Keyword:** BROYDEN
$$0, 7, 0.8, 1000. = \text{IBROYD, N\_ITER, ALPHAM, BROTRI}$$
For IBROYD=1, the Broyden method is used to accelerate the convergence, see Section 2.8. For N_ITER=0, variables N_ITER, ALPHAM, and BROTRI, which are read on this line, are ignored, that is, the previous values are kept. N_ITER is the number of iterations used to approximate the inverse Jacobian, ALPHAM is the value of the parameter $\alpha$ of the linear mixing to which is added the Broyden correction, and BROTRI triggers an automatic switch to the Broyden method, that is, when the absolute value of the stability energy becomes lower than BROTRI, iterations are changed to the Broyden scheme. A large value of BROTRI ensures that the Broyden method is used from the very first iteration. IBROYD=1 is incompatible with I_YUKA>0, ICOUDI=2, or ICOUEX=2. The Broyden method is implemented only in the FORTRAN-90 version of HFODD.

## 3.5 Constraints

**Keyword:** MULTCONSCA
$$1, 0., 0., 0 = \text{LAMBDA, STIFFG, GASKED, IFLAGG}$$

For IFLAGG=1, the total scalar multipole moment of the given multipolarity $\lambda$ is constrained. Values of LAMBDA, STIFFG, and GASKED correspond, respectively, to $\lambda$, $C_\lambda$, and $\bar{Q}_\lambda$ in Eq. (80). For IFLAGG=0, there is no scalar constraint in the given multipolarity. The constrained multipolarity LAMBDA cannot be equal to 0 or exceed NMUCON. For conserved parity IPARTY=1, only even moments can be constrained.

**Keyword:** NORBCONSTR
$$0 = \text{NO\_ORB}$$
For NO_ORB=1, constraints on the intrinsic spin only are used and the orbital part of the angular momentum is not constrained. NO_ORB=1 requires IROTAT=1.

**Keyword:** BOHR_BETAS
$$4, 0, 1 = \text{NEXBET, IPRIBE, IPRIBL}$$
For a given set of electric multipole moments, the code calculates and prints the corresponding first-order (for IPRIBL=1) and/or exact (for IPRIBE=1) Bohr deformation parameters, see Section 2.5. For IPRIBE=1, in case the code fails to find the exact values, the first-order values are printed irrespective of the value of IPRIBL. The exact values are sought for multipolarities up to NEXBET, which must not be greater than NMUPRI. The approximate values are printed up to multipolarity of NMUPRI.

**Keyword:** SCHIFF_MOM
$$0 = \text{ISCHIF}$$

For ISCHIF=1, the surface moments are everywhere in the code replaced by the Schiff moments, see Section 2.6.

## 3.6 Output-file parameters

**Keyword:** HFBMEANFLD
$$0 = \text{IMFHFB}$$
For IMFHFB=1, eigenvalues of the HFB mean-field s.p. Hamiltonian or Routhian are printed. IMFHFB=1 requires IPAHFB=1.

**Keyword:** PRINT_AMP
$$0, 999, 1, 1, 1, 0, 0$$
$$\text{ISLPRI, ISUPRI, IENPRI, ISRPRI, IMIPRI, IKEPRI, IRMPRI}$$

These parameters govern the printing of the AMP results. If not restricted by the values of the doubled angular momenta for which the calculations are performed, IPROMI and IPROMA, the results are printed only for the doubled angular momenta between ISLPRI and ISUPRI. For IENPRI=1, the AMP energies are printed. In addition, for IENPRI=2, the AMP kernels are also printed. For ISRPRI=1, the sum rules are printed and compared with the HF average values. For IMIPRI=1, the energies of the $K$-mixed states are printed. For IKEPRI=1 and/or IRMPRI=1, the proton reduced kernels and/or reduced matrix elements, respectively, are printed.

**Keyword:** TRANCUTPRI
$$0., 0., 0., = \text{QMUCUT, QMACUT, QSICUT}$$
Values of the electric, magnetic, and surface or Schiff proton kernels and reduced matrix elements are printed only if their absolute values are larger than, respectively, QMUCUT, QMACUT, QSICUT. This may avoid printing long lists of very small or zero values.

**Keyword:** ONE_LINE
$$1 = \text{I1LINE}$$
For I1LINE$\neq 0$, a one-line convergence report is printed at each iteration. For I1LINE=1, the code prints the values of deformation $\gamma$, total angular momentum, total angular frequency $\omega$, and angle between vectors of angular momentum and frequency. For I1LINE=2, the code prints the values of neutron and proton pairing gaps and Lipkin-Nogami parameters $\lambda_2$.

**Keyword:** PRINT_VIOL
$$0 = \text{IVIPRI}$$
For IVIPRI=1, the code prints integrals of several symmetry-violating terms. IVIPRI=1 requires IROTAT=1.

**Keyword:** NILSSONLAB
$$3 = \text{NILXYZ}$$
This option used for NILXYZ=1, 2, or 3, allows for printing the Nilsson labels defined with respect to the $x$, $y$, or $z$ axis, respectively. This feature is useful when the symmetry axis of a nucleus and its spin are aligned along the $x$ or $y$ axis and not along the $z$ axis, for which the standard Nilsson labels are defined. This is particularly important for analyzing configurations of band heads of odd nuclei within the conserved $y$-signature-symmetry limit.

## 3.7 Files

**Keyword: WAVEF_FILE**
$$\text{HFODD.WFN} = \texttt{FILWAV}$$
`CHARACTER*68` file name of the wave function file. Must start at the 13-th column of the data line. The binary wave function files must exist if `IPRGCM=2`, and will be read, see the keyword `PROJECTGCM`.

**Keyword: KERNELFILE**
$$\text{HFODD.KER} = \texttt{FILKER}$$
`CHARACTER*68` file name of the kernel file. Must start at the 13-th column of the data line. The binary kernel files are written and read if `ISAKER=1`, see the keyword `SAVEKERNEL`.

**Keyword: YUKAWASAVE**
$$1 = \texttt{IWRIYU}$$
For `IWRIYU=1` and `I_YUKA`$\geq$`2`, the Yukawa file is saved on disc after each iteration is completed. The file contains the matrix elements of the Yukawa mean field. For `IWRIYU=0` and `I_YUKA`$\geq$`2`, the file is saved only once, after all iterations are completed. For `IWRIYU=−1`, the file is never saved. The Yukawa calculations can be restarted by using the Yukawa file. Therefore, `IYCONT=1` requires setting `IWRIYU=0` or 1 in the run that is to be continued. Equivalently, the Yukawa calculations can be restarted by using the fields file, which requires setting `IWRIFI=0` or 1 in the run that is to be continued.

**Keyword: REPYUKFILE**
$$\text{HFODD.YUP} = \texttt{FILYUP}$$

`CHARACTER*68` file name of the Yukawa file. Must start at the 13-th column of the data line. The binary Yukawa file with the name defined in `FILREP` must exist if `IYCONT=1`, and will be read. If the filenames `FILREP` and `FILREC` are identical, the Yukawa file will be subsequently overwritten as a new Yukawa file.

**Keyword: RECYUKFILE**
$$\text{HFODD.YUC} = \texttt{FILYUC}$$

`CHARACTER*68` file name of the Yukawa file. Must start at the 13-th column of the data line. If `IWRIYU=1`, binary Yukawa file is written after each HF iteration. It contains complete information that allows restarting the Yukawa calculations in another run of the code. To restart, one has to specify `IYCONT=1` and provide the name of the file by defining `FILREP`.

**Keyword: LIPKINSAVE**
$$1 = \texttt{IWRILI}$$
For `IWRILI=1` and `LIPKIN=1` or `LIPKIP=1`, the Lipkin-Nogami file is saved on disc after each iteration is completed. The file contains the matrix elements of the particle density matrices. For `IWRILI=0` and `LIPKIN=1` or `LIPKIP=1`, the file is saved only once, after all iterations are completed. For `IWRILI=−1`, the file is never saved. The Lipkin-Nogami calculations can be restarted by using the Lipkin-Nogami file. Therefore, `ILCONT=1` requires setting `IWRILI=0` or 1 in the run that is to be continued. Equivalently, the Lipkin-Nogami calculations can be restarted by using the fields file, which requires setting `IWRIFI=0` or 1 in the run that is to be continued.

**Keyword:** REPLIPFILE

$$\text{HFODD.LIP} = \texttt{FILLIP}$$

`CHARACTER*68` file name of the Lipkin-Nogami file. Must start at the 13-th column of the data line. The binary Lipkin-Nogami file with the name defined in `FILLIP` must exist if `ILCONT=1`, and will be read. If the filenames `FILLIP` and `FILLIC` are identical, the Lipkin-Nogami file will be subsequently overwritten as a new Lipkin-Nogami file.

**Keyword:** RECLIPFILE

$$\text{HFODD.LIC} = \texttt{FILLIC}$$

`CHARACTER*68` file name of the Lipkin-Nogami file. Must start at the 13-th column of the data line. If `IWRILI=1`, binary Lipkin-Nogami file is written after each HF iteration. It contains complete information that allows restarting the Lipkin-Nogami calculations in another run of the code. To restart, one has to specify `ILCONT=1` and provide the name of the file by defining `FILLIP`.

**Keyword:** FIELD_SAVE

$$-1 = \texttt{IWRIFI}$$

For `IWRIFI=1`, the field file, is saved on disc after each iteration is completed. The file contains the matrix elements of the mean field. For `IWRIFI=0`, the file is saved only once, after all iterations are completed. For `IWRIFI=−1`, the file is never saved. To restart calculations of the exact Coulomb exchange energy, the field file must exist. `IFCONT=1` requires setting `IWRIFI=0` or 1 in the run that is to be continued.

**Keyword:** FIELD_OLD

$$0 = \texttt{IWRIOL}$$

For `IWRIOL=1`, in the last iteration the mean fields are not updated, that is, the mixing fractions $\epsilon$ are set equal to 1, irrespective of values of the `SLOWEV`, `SLOWOD` and `SLOWPA` parameters. In this way, information stored on the record file corresponds to the last but one iteration. Only then, restarting of the calculation from the field file leads to a smooth continuation of iterations.

**Keyword:** REP_FIELDS

$$\text{HFODD.FIP} = \texttt{FILFIP}$$

`CHARACTER*68` file name of the field file. Must start at the 13-th column of the data line. The binary field file with the name defined in `FILFIP` must exist if `IFCONT=1`, and will be read. If the filenames `FILFIP` and `FILFIC` are identical, the field file will be subsequently overwritten as a new field file.

**Keyword:** REC_FIELDS

$$\text{HFODD.FIC} = \texttt{FILFIC}$$

`CHARACTER*68` file name of the field file. Must start at the 13-th column of the data line. If `IWRIFI=1`, binary field file is written after each HF iteration. It contains complete information that allows restarting the calculations from the matrix elements of fields. To restart, one has to specify `IFCONT=1` and provide the name of the file by defining `FILFIP`.

## 3.8 Starting the iteration

**Keyword: CONTYUKAWA**
$$0 = \text{IYCONT}$$

For IYCONT=1, results stored in the Yukawa file are used to define the Yukawa fields in the first iteration; otherwise the Yukawa fields are set equal to zero. When the Yukawa fields are taken into account (i.e., for I_YUKA=2), and if a smooth restart and continuation of iterations from previously stored results is required, value of IYCONT=1 must be used. IYCONT=1 is incompatible with either of I_YUKA<2 or ICONTI=0.

**Keyword: CONTLIPKIN**
$$0 = \text{ILCONT}$$

For ILCONT=1, results stored in the Lipkin-Nogami file are used to define the density matrix required for the Lipkin-Nogami calculations in the first iteration; otherwise the required density matrix is set equal to zero. For the Lipkin-Nogami calculations (i.e., for LIPKIN=1 or LIPKIP=1), and if a smooth restart and continuation of iterations from previously stored results is required, value of ILCONT=1 must be used. ILCONT=1 is incompatible with either of LIPKIN=LIPKIP=0 or ICONTI=0.

**Keyword: CONTFIELDS**
$$0 = \text{IFCONT}$$

For IFCONT=1, results stored in the field file are used to define the matrix elements of fields in the first iteration; otherwise the matrix elements are recalculated. For the Gaussian-expansion method used to calculate the Coulomb energy and Coulomb mean field, that is for ICOUDI=2 or ICOUEX=2, and if a smooth restart and continuation of iterations from previously stored results is required, value of IFCONT=1 must be used. IFCONT=1 is incompatible with ICONTI=0.

# 4 Output file

Together with the FORTRAN source code in the file `hfodd.f`, two examples of the output file are provided in `ge064-a.out` and `sn120-b.out`. Selected lines from the file `ge064-a.out` are presented in section TEST RUN OUTPUT below. This output file corresponds to the input file `ge064-a.dat`, reproduced in section TEST RUN INPUT below.

The output file `ge064-a.out` contains the following new sections:

Section CODE COMPILED WITH THE... lists the values of the most important array dimensions and switches declared in the PARAMETER statements, fixed at the compilation stage.

Sections BOHR DEFORMATIONS (EXACT MULTIPOLE MOMENTS) give values of the Bohr deformation parameters, determined as described in Section 2.5.

Section KERNELS AND AVERAGE VALUES... gives values of the diagonal norm kernels $N_{KK}$ (5) and average AMP energies $H_{KK}/N_{KK}$ for different values of $I$ and $K$. Real and imaginary parts of $N_{KK}$ are both printed, although all imaginary parts should be equal to zero when the numerical precision is sufficiently good.

Section `SUM RULES`... compares real and imaginary parts of the sum rules (49) and (50) with the corresponding HF average values. Results are printed for the norm kernels $N_{KK}^I$, which must sum up to 1. This condition is a primary test of whether a sufficient number of the angular momenta are included in the sum of Eq. (48). In the example of the file `ge064-a.out`, the sum rule for the norm kernel `NORM= 0.3248` indicates a much too small value of $I_{max}$ and too small numbers of the G-T and G-L integration nodes.

Sections `REDUCED KERNELS`... give values of the reduced kernels (56) of electric and magnetic transition operators.

Sections `RESULTS OF THE K-MIXING`... give values of the energies $E_i$ of $K$-mixed states, see Eq. (3) and values of norm eigenvalues $n_m$, see Eq. (9).

Sections `REDUCED MAT.ELEMS`... give values of the reduced matrix elements (57) of electric and magnetic transition operators calculated for the $K$-mixed states.

Section `NUMBERS OF CALLS TO SUBROUTINES` gives the statistics of calls to subroutine, which together with the section `EXECUTION TIMES IN SUBROUTINES` illustrates the work flow of the code.

## 5 FORTRAN source file

The FORTRAN source code is provided in the file `hfodd.f` and can be modified in several places, as described in this section.

### 5.1 FORTRAN-90 version

Similarly as for the previous version (v2.08k), the code HFODD version (v2.38j) is written in FORTRAN-77 and FORTRAN-90. In the FORTRAN source code provided in the file `hfodd.f`, all the FORTRAN-90 features are commented out and inactive. However, very simple modifications of the source code can easily be performed to transform the code HFODD to FORTRAN-90, as described in Section (IV-5.2).

The present version of the code HFODD version (v2.38j) is the last one that works under FORTRAN-77; future releases will only use FORTRAN-90 programming.

A set of c-shell and ex-editor scripts is provided within the HFODD distribution file, which allows an easy installation, compilation, and execution of the FORTRAN-90 version on a Linux computer running the INTEL© FORTRAN COMPILER.

### 5.2 Library subroutines

*5.2.1 BLAS.*
The code HFODD requires an implementation of the BLAS (Basic Linear Algebra Subprogams) interface for common dense vector and matrix computations. A reference implementation of these subroutines can be downloaded from
http://www.netlib.org/blas/blas.tgz,
but for peak performance it is recommended that an optimized version of these routines should be installed. Optimized machine-specific BLAS (such as Sun Performance Library and Intel Math Kernel Library) are available from most hardware vendors, and there are also third-party optimized implementations such as GOTO and ATLAS available for various architectures.

The BLAS subroutines are in the `REAL*8/COMPLEX*16` version, and should be compiled without promoting real numbers to the double precision. On the other hand, the code HFODD itself does require compilation with an option promoting to double precision. Therefore, the code and the BLAS package should be compiled separately, and then should be linked together.

*5.2.2 Diagonalization subroutines.*

The code HFODD requires an external subroutine that diagonalizes complex hermitian matrices. Version (v1.60r) (see II) has been prepared with an interface to the NAGLIB subroutine F02AXE, version (v1.75r) (see III) with an interface to the LAPACK subroutine ZHPEV, and version (v2.08i) (see IV) with an interface to the LAPACK subroutine ZHPEVX. In the present version (v2.38j), all these interfaces remain supported and can be activated as described in II–IV. However, the recommended interface is now to the LAPACK subroutine ZHEEVR, as described in this section.

In version (v2.38j) we have implemented interface to the LAPACK subroutine ZHEEVR, which can be downloaded (with dependencies) from
`http://www.netlib.org/cgi-bin/netlibfiles.pl?filename=/lapack/complex16/zheevr.f`
This subroutine works with unpacked matrices and hence performs calculations in less CPU time at the expense of using a larger memory. Alternatively, the entire LAPACK library can be downloaded from
`http://www.netlib.org/lapack/lapack.tgz`

Subroutine ZHEEVR and its dependencies are in the `REAL*8/COMPLEX*16` version, and should be compiled without promoting real numbers to the double precision. Therefore, the code and the ZHEEVR package should be compiled separately, and then should be linked together.

In order to activate the interface to the LAPACK ZHEEVR subroutine, the following modifications of the code HFODD (v2.38j) have to be made:

1. Change everywhere the value of parameter `I_CRAY=1` into `I_CRAY=0`.

2. Change everywhere the value of parameter `IZHPEV=0` into `IZHPEV=3`.

3. If your compiler and linker do not support undefined externals, or subroutines called with different parameters, remove calls to subroutines CGEMM, F02AXE, ZHPEV, and ZHPEVX.

*5.2.3 Matrix inversion subroutines.*

The code HFODD requires external subroutines that invert complex matrices and calculate their determinants. In version (v2.38j) we have implemented interfaces to the LINPACK subroutines ZGEDI and ZGECO, which can be downloaded from
http://www.netlib.org/linpack/zgedi.f , and
http://www.netlib.org/linpack/zgeco.f , respectively, together with
http://www.netlib.org/linpack/zgefa.f

These subroutines are in the `REAL*8/COMPLEX*16` version, and should be compiled without promoting real numbers to the double precision. Therefore, the code and these subroutines should be compiled separately, and then should be linked together.

# 6 Acknowledgments

This work was supported in part by the Polish Ministry of Science and Higher Education under Contract No. N N202 328234, by the Academy of Finland and University of Jyväskylä within the FIDIPRO programme, by the UNEDF SciDAC Collaboration under the U.S. Department of Energy grant No. DE-FC02-07ER41457, and by the computational grant from the Interdisciplinary Centre for Mathematical and Computational Modeling (ICM) of the University of Warsaw.

# TEST RUN INPUT

```
================================================================================
| This file (ge064-a.dat) contains input data for the code HFODD (v2.38j)      |
================================================================================
                        ---------- General data   ----------
NUCLIDE
             32    32
ITERATIONS
             20
ITERAT_EPS
             0.000001
MAXANTIOSC
             5
PING-PONG
             0.0   3
CHAOTIC
             0
PHASESPACE
             0     0      0     0
                        ---------- Interaction   -----------
SKYRME-SET
             SIII
SKYRME-STD
             1     1      0     0    0
                        ---------- Symmetries    -----------
SIMPLEXY
             1
SIGNATUREY
             1
PARITY
             -1
ROTATION
             0
TSIMPLEX3D
             1     1      1
PAIRING
             0
HFB
             0
                        ---------- Configurations --------
VACSIG_NEU           PPSP PPSM PMSP PMSM
                      7    7    9    9
VACSIG_PRO           PPSP PPSM PMSP PMSM
                      7    7    9    9
OPTI_GAUSS
             1
                        --- Parameters of the HO basis  ---
BASIS_SIZE
             14   680   800.
SURFAC_PAR
             32    32   1.23
SURFAC_DEF
             2     0    0.00
SURFAC_DEF
             4     0    0.00
                        ---------- Constraints   -----------
OMEGAY
             0.00
MULTCONSTR
             2     0    1.0    2.7   1
MULTCONSTR
             2     2    1.0   -1.3   1
                        ---- Output-file  parameters  -----
PRINT-ITER
             1     0      1
PRINT-MOME
             0     0      1
PRINT-INTR
             0
EALLMINMAX
             -40.  0.
EQUASI_MAX
             10.0
MAX_MULTIP
             2     4      6
BOHR_BETAS
             6     1      1
                        ------------ Files  --------------
REVIEW
             0
RECORDFILE
             ge064-a.rec
RECORDSAVE
             1
REPLAYFILE
             ge064-a.rec
COULOMFILE
```

```
COULOMSAVE  ge064-a.cou
            1        1
                         -----  Starting the iteration  -----
RESTART
            0
                         ----------  Calculate  -------------
EXECUTE
                         ----------  Next Run   -------------
ITERATIONS
            5000
MULTCONSTR
            2        0    1.0      2.7   0
MULTCONSTR
            2        2    1.0     -1.3   0
BROYDEN
            1        0    0.0      0.0
RESTART
            1
EXECUTE
                         ----------  Next Run   -------------
ROTATION
            1
TSIMPLEX3D
            1       -1     1
OMEGAY
            0.575
EXECUTE
                         ----------  Next Run   -------------
ITERATIONS
            1
SIMPLEXY
            0
SIGNATUREY
            0
TSIMPLEX3D
            0        0     0
PROJECTGCM
            1   0   10    10    10   0     1    1    0
SAVEKERNEL
            1
PRINT_AMP
            0   10    1     1     1   1     1
TRANCUTPRI
            0.001    0.001    0.001
CUTOVERLAP
            1        0.01     0.00
RECORDSAVE
           -1
KERNELFILE
            ge064-a.ker
BROYDEN
            0        0    0.0      0.0
EXECUTE
                         ----------  Terminate  -------------
ALL_DONE
```

# TEST RUN OUTPUT

```
*****************************************************************************
*                                                                           *
*   HFODD    HFODD    HFODD    HFODD    HFODD    HFODD    HFODD    HFODD    *
*                                                                           *
*****************************************************************************
*                                                                           *
*           SKYRME-HARTREE-FOCK-BOGOLYUBOV CODE VERSION: 2.38J              *
*                                                                           *
*           NO SYMMETRY-PLANES AND NO TIME-REVERSAL SYMMETRY                *
*                                                                           *
*            DEFORMED CARTESIAN HARMONIC-OSCILLATOR BASIS                   *
*                                                                           *
*****************************************************************************
*                                                                           *
*             J. DOBACZEWSKI, B.G. CARLSSON, J. DUDEK, J. ENGEL             *
*           P. OLBRATOWSKI, P. POWALOWSKI, M. SADZIAK, W. SATULA            *
*           N. SCHUNCK, A. STASZCZAK, M. STOITSOV, M. ZALEWSKI              *
*                              AND H. ZDUNCZUK                              *
*                                                                           *
*                 INSTYTUT FIZYKI TEORETYCZNEJ, WARSZAWA                    *
*                                                                           *
*                                 1993-2009                                 *
*                                                                           *
*****************************************************************************
*****************************************************************************
*                                                                           *
*  CODE COMPILED WITH THE FOLLOWING ARRAY DIMENSIONS AND SWITCHES:          *
*                                                                           *
*****************************************************************************
*                                                                           *
*  NDBASE =  680  NDSTAT =  181  NDXHRM =   30  NDYHRM =   30  NDZHRM =  30 *
*                                                                           *
*  NDMAIN =   14  NDMULT =    9  NDMULR =    4  NDLAMB =    9  NDITER = 5000*
*                                                                           *
*  NDAKNO =   50  NDBKNO =   50  NDPROI =   50  NDCOUL =   80  NDPOLS =  25 *
*                                                                           *
*  IPARAL =    0  I_CRAY =    0  IZHPEV =    3                              *
*                                                                           *
*****************************************************************************
*                                                                           *
* REDUCED KERNELS AND MATRIX ELEMENTS:                                      *
* FOR ELECTRIC MULTIPOLES CALCULATED UP TO = 2, PRINTED IF LARGER THAN 0.001*
* FOR MAGNETIC MULTIPOLES CALCULATED UP TO = 1, PRINTED IF LARGER THAN 0.001*
* FOR SURFACE  MULTIPOLES CALCULATED UP TO = 0, PRINTED IF LARGER THAN 0.001*
*                                                                           *
*****************************************************************************
*                                                                           *
* PRINTING THE RESULTS FOR ANGULAR-MOMENTUM-PROJECTED STATES:  0 < 2*I <  10*
*                            AVERAGE ENERGIES OF PROJECTED STATES: YES      *
*                            KERNELS BETWEEN THE PROJECTED STATES:  NO      *
*                            SUM RULES COMPARED TO THE H-F VALUES: YES      *
*                            K-MIXED ENERGIES OF PROJECTED STATES: YES      *
*             REDUCED  K E R N E L S  BETWEEN THE PROJECTED STATES: YES     *
*             REDUCED MATRIX ELEMENTS BETWEEN THE  K-MIXED  STATES: YES     *
*                                                                           *
*****************************************************************************
*                                                                           *
*  BOHR DEFORMATIONS   (EXACT MULTIPOLE MOMENTS)                     TOTAL  *
*                                                                           *
*****************************************************************************
*                                                                           *
*  B10 =    ZERO  B11 =    ZERO  ............  ............  ............  *
*                                                                           *
*  B20 =  0.2094  B21 =    ZERO  B22 = -0.1138  ............  ............  *
*                                                                           *
*  B30 =    ZERO  B31 =    ZERO  B32 =    ZERO  B33 =    ZERO  ............ *
*                                                                           *
*  B40 = -0.0306  B41 =    ZERO  B42 = -0.0348  B43 =    ZERO  B44 = -0.0049*
*                                                                           *
*  B50 =    ZERO  B51 =    ZERO  B52 =    ZERO  B53 =    ZERO  B54 =    ZERO*
*                                                              B55 =    ZERO*
*                                                                           *
*  B60 = -0.0054  B61 =    ZERO  B62 =  0.0147  B63 =    ZERO  B64 = -0.0098*
*                                                              B65 =    ZERO  B66 =-7.7E-04*
*                                                                           *
*****************************************************************************
*                                                                           *
*  KERNELS AND AVERAGE VALUES OBTAINED FOR ANGULAR-MOMENTUM PROJECTED STATES*
*                                                                           *
*****************************************************************************
*                                                                           *
*                       REAL(KERNEL)    IMAG(KERNEL)   AVERAGE ENERGY       *
*                                                                           *
*  I,K=   0    0   NORM = 0.008035212882  0.000000000000    -542.137728     *
*  I,K=   1   -1   NORM = 0.000867254021  0.000000000000    -509.681812     *
*  I,K=   1    0   NORM = 0.000000000000  0.000000000000    -967.529412     *
*  I,K=   1    1   NORM = 0.000867254021  0.000000000000    -509.681812     *
*  I,K=   2   -2   NORM = 0.014229645333  0.000000000000    -541.456596     *
*  I,K=   2   -1   NORM = 0.003145537421  0.000000000000    -530.120652     *
```

```
*  I,K=    2    0    NORM =   0.034483057875   0.000000000000           -541.966975  *
*  I,K=    2    1    NORM =   0.003145537421   0.000000000000           -530.120652  *
*  I,K=    2    2    NORM =   0.014229645333   0.000000000000           -541.456596  *
**********************************************************************************
*                                                                                  *
*  SUM RULES OBTAINED FOR ANGULAR-MOMENTUM PROJECTED STATES VS THE HF VALUES       *
*                                                                                  *
**********************************************************************************
*                                                                                  *
*                            REAL(SUM RULE)   IMAG(SUM RULE)         HF VALUE      *
*                                                                                  *
*                    NORM=    0.324825071083   0.000000000000                      *
*                  SKYRME=    -587.075736         0.000000        -1808.113218     *
*                  EKIN_T=     358.576775         0.000000         1105.378182     *
*                  COUL_D=      57.290170         0.000000          176.353367     *
*                  COUL_E=      -4.008437         0.000000          -12.345178     *
*                                                                                  *
*  MULTIPOLE L= 0 M= 0 QMUL_P=       10.394402         0.000000         32.000000  *
*           L= 1 M=-1 QMUL_P=        0.000000         0.000000          0.000000   *
*           L= 1 M= 0 QMUL_P=        0.000000         0.000000          0.000000   *
*           L= 1 M= 1 QMUL_P=        0.000000         0.000000          0.000000   *
*           L= 2 M=-2 QMUL_P=       -0.206203         0.000000          0.000000   *
*           L= 2 M=-1 QMUL_P=        0.000000         0.032421          0.000000   *
*           L= 2 M= 0 QMUL_P=        0.432164         0.000000          1.333826   *
*           L= 2 M= 1 QMUL_P=        0.000000        -0.032421          0.000000   *
*           L= 2 M= 2 QMUL_P=       -0.206203         0.000000         -0.621629   *
*                                                                                  *
**********************************************************************************
*                                                                                  *
*  REDUCED  KERNELS  OF PROTON MULTIPOLE MOMENTS [UNITS:   (10 FERMI)^LAMBDA]      *
*  ONLY THE RESULTS WITH ABSOLUTE VALUES LARGER THAN 0.001 ARE BELOW PRINTED       *
*                                                                                  *
**********************************************************************************
*                                         *                                        *
*  IL   KL    L   IR    KR     <||Q||>    *  IL   KL    L   IR    KR     <||Q||>   *
*                                         *                                        *
*   0    0    2    2    -2    -0.008      *   2   -2    2    2    -2     0.025     *
*   2    0    2    2    -2    -0.017      *   2    2    2    2    -2     0.025     *
*   1   -1    2    2    -1     0.003      *   1    1    2    2    -1     0.003     *
*   0    0    2    2     0     0.024      *   2   -2    2    2     0    -0.017     *
*   2    0    2    2     0    -0.055      *   2    2    2    2     0    -0.017     *
*   1   -1    2    2     1    -0.003      *   1    1    2    2     1    -0.003     *
*   0    0    2    2     2    -0.008      *   2   -2    2    2     2     0.025     *
*   2    0    2    2     2    -0.017      *   2    2    2    2     2     0.025     *
**********************************************************************************
*                                                                                  *
*  RESULTS OF THE K-MIXING CALCULATION FOR THE FOLLOWING OVERLAP CUTOFF DATA       *
*  ICUTOV=1, CUTOVE= 1.0E-02, CUTOVF= 0.0E+00                                      *
*                                                                                  *
**********************************************************************************
*                                         |                                        *
*   I    N    OVERLAP       ENERGY        |   I    N    OVERLAP       ENERGY      *
*                                         |                                        *
*-----------------------------------------|----------------------------------------*
*   2    1    4.02E-02    -541.817486     |   3    1    3.78E-02    -540.997225   *
*   2    2    2.71E-02    -541.344808     |                                        *
*-----------------------------------------|----------------------------------------*
*   4    1    7.90E-02    -541.775088     |   5    1    5.06E-02    -540.367910   *
*   4    2    3.04E-02    -540.196719     |   5    2    2.04E-02    -537.665199   *
*   4    3    2.06E-02    -538.486590     |                                        *
*-----------------------------------------|----------------------------------------*
**********************************************************************************
*                                                                                  *
*  REDUCED MAT.ELEMS OF PROTON MULTIPOLE MOMENTS [UNITS:   (10 FERMI)^LAMBDA]      *
*  ONLY THE RESULTS WITH ABSOLUTE VALUES LARGER THAN 0.001 ARE BELOW PRINTED       *
*                                                                                  *
**********************************************************************************
*                                         *                                        *
*  IL   NL    L   IR    NR     <||Q||>    *  IL   NL    L   IR    NR     <||Q||>   *
*                                         *                                        *
*   2    1    2    2     1    -1.142      *   2    2    2    2     1    -1.419     *
*   2    1    2    2     2    -1.419      *   2    2    2    2     2     1.142     *
*   3    1    2    3     1    -0.007      *   2    1    2    4     1    -2.353     *
*   2    2    2    4     1     0.543      *   4    1    2    4     1    -0.518     *
*   4    2    2    4     1    -1.023      *   4    3    2    4     1    -0.417     *
*   2    1    2    4     2    -0.554      *   2    2    2    4     2    -1.264     *
*   4    1    2    4     2    -1.023      *   4    2    2    4     2    -2.302     *
*   4    3    2    4     2     0.763      *   2    1    2    4     3     0.159     *
*   2    2    2    4     3    -0.831      *   4    1    2    4     3    -0.417     *
*   4    2    2    4     3     0.763      *   4    3    2    4     3     2.832     *
*   3    1    2    5     1     2.170      *   5    1    2    5     1    -0.937     *
*   5    2    2    5     1     1.296      *   3    1    2    5     2     0.381     *
*   5    1    2    5     2     1.296      *   5    2    2    5     2     0.982     *
**********************************************************************************
*                                                                                  *
*  NUMBERS OF CALLS TO SUBROUTINES                                                 *
*                                                                                  *
```

```
*******************************************************************************
*                                                                             *
*     26730 => INTMUL     4002 => INTSPI     4001 => INTORB     2354 => BEGINT *
*                                                                             *
*      2324 => INTKIN     2162 => DENSHF     2162 => DENMAC     2000 => ROTWAV *
*                                                                             *
*      1162 => MOMETS     1162 => MOMSIF     1162 => MAGMOM     1084 => COUMAT *
*                                                                             *
*      1084 => BEGINC     1084 => INTCOU      546 => INTSOR      541 => DIAMAT *
*                                                                             *
*       358 => INTMAS      358 => INTCEN      249 => ROTMOM      162 => AVPARI *
*                                                                             *
*       162 => MOMVMU      162 => SPIMOM      162 => LINAVR      160 => DIASIG *
*                                                                             *
*       160 => INTEGH      106 => SPAVER       83 => RECORD       81 => SKFILD *
*                                                                             *
*        81 => SHICOE       81 => SHIMOM       81 => SHISIF       81 => SHIMAG *
*                                                                             *
*        59 => DOBROY       14 => AVANGY       14 => AVOBSE       14 => AVSIMP *
*                                                                             *
*        14 => NILABS        5 => RECOUL        2 => NILSON        2 => DIASIZ *
*                                                                             *
*         1 => HFODD         1 => POWALL        1 => PROANG        1 => DIAPRO *
*                                                                             *
*         1 => REDPRO                                                          *
*                                                                             *
*******************************************************************************
```